\documentstyle[pre,preprint,aps,epsfig,amsmath]{revtex}

\newcommand{\beq}{\begin{equation}}
\newcommand{\eeq}{\end{equation}}
\newcommand{\bea}{\begin{eqnarray}}
\newcommand{\eea}{\end{eqnarray}}
\newcommand{\barr}{\begin{array}}
\newcommand{\earr}{\end{array}}

\newcommand{\bm}{\mathbf}
\newcommand{\pdev}[2]{\frac{\partial #1}{\partial #2}}
\begin{document}
%\twocolumn[\hsize\textwidth\columnwidth\hsize\csname 
%@twocolumnfalse\endcsname

\date{\today}
\title{ A kinetic approach to granular gases }

\author{
       A. Puglisi$^1$, 
       V. Loreto$^{2,3,1}$,
        U. Marini Bettolo Marconi$^{4,5}$
        and A. Vulpiani$^{1}$}

\address{(1) Dipartimento di Fisica, Universit\`a La Sapienza,
Piazzale A. Moro 2, 00185 Roma, Italy and
 Istituto Nazionale di Fisica della Materia, Unit\`a di Roma}         

\address{(2) P.M.M.H., Ecole Sup\'erieure de Physique et Chimie
  Industrielles, 10, rue Vauquelin, 75231 Paris, France}

\address{(3) ENEA Research Center, Localit\`a Granatello C.P. 32 80055
  Portici, Napoli, Italy}

\address{(4) Dipartimento di Matematica e Fisica, Universit\`a di Camerino,
Via Madonna delle Carceri, I-62032 , Camerino, Italy and
Istituto Nazionale di Fisica della Materia, Unit\`a di Camerino} 

\address{(5) Istituto Nazionale di Fisica Nucleare, Sezione di Perugia} 

\maketitle

\begin{abstract}
 We address the problem of the so-called ``granular gases'',
 i.e. gases of massive particles in rapid movement undergoing
 inelastic collisions. We introduce a class of models of driven granular gases
 for which the stationary state is the
 result of the balance between the
 dissipation and the random forces which inject energies. 
 These models exhibit a
 genuine thermodynamic limit, i.e. at fixed density the mean values of
 kinetic energy and dissipated energy per particle are independent of
 the number $N$ of particles, for large values of $N$. One
 has two 
 regimes: when the typical relaxation  time $\tau$ of the 
 driving Brownian process is small compared with the mean collision time $\tau_c$
the spatial density is nearly homogeneous and the velocity probability
distribution is gaussian.
In the opposite limit $\tau \gg \tau_c$ one has strong spatial
clustering, with a fractal distribution of particles,
and the  velocity probability  distribution strongly 
deviates from the gaussian one. 
Simulations performed in one and two dimensions under the 
{\it Stosszahlansatz} Boltzmann  approximation
confirm the scenario. Furthermore we analyze
the instabilities bringing to the spatial and the velocity
clusterization. 
Firstly, in the framework of a mean-field model, we explain
how the existence of the inelasticity can bring to a spatial clusterization; on
the other side we discuss, in the framework of a Langevin dynamics
treating the collisions in a mean-field way, how a
non-gaussian distribution of velocity can arise. The comparison
between the numerical and the analytical results exhibits an excellent
agreement.

\end{abstract}
\vspace{0.2cm}
\vskip2pc

\section{Introduction: hydrodynamics and gases}

Granular systems (sand, powders, seeds, cements, etc.) have been extensively 
studied, in the last two decades, by
means of analytical investigations, experiments and computer simulations.
The rich and intriguing  phenomenology is well known to engineers and 
the need of a better comprehension of granular behaviors is widely 
recognized in applied sciences as well as in theoretical physics.
A quite exhaustive review may be found in~\cite{jaeger}. Problems in granular 
systems are roughly divided in quasi-static (sand piles, distribution of static
forces, compaction, fractures propagation, etc.) and dynamical ones 
(all kind of
flows, convection and segregation, pattern formation, fluidized beds, etc.). 
In the latter class, large collections of inelastic particles
are involved in fluid-like rapid dynamics, therefore  the hydrodynamics
approach seems to be the natural one. The main granular hydrodynamics 
theories are reviewed in~\cite {campbell}: they are all based  on 
non-equilibrium conservation laws~\cite{mazur} for mass, momentum and energy:
\beq \frac{d \rho}{dt}+ \rho \nabla \cdot \mathbf{u}=0 \end{equation}
\beq \rho \frac{d {\bm u}}{dt}=-\nabla \cdot \bm{\hat{\tau}}+\rho \bm{g}
\end{equation}
\beq \frac{1}{2} \rho \frac{dT}{dt}=-\nabla \cdot \bm{q}+ \bm{\hat{\tau}}: 
\nabla
\bm{u}-\Gamma \end{equation}
where $\rho$ is the local density, $\bm{u}$ is the local velocity vector, 
$d/dt=\partial/\partial t+\bm{u} \cdot \nabla$ is the Lagrangian derivative,
$\hat{\tau}$ is the total stress tensor 
($\hat{\tau}=\hat{\tau}_s+\hat{\tau}_c$, as transport of 
momentum has two contributions: a ``streaming'' term and a 
``collisional'' one), $\bm{g}$
is the body-force vector (gravity or else), $T$ is the so-called \emph{
granular temperature} $T=<(\bm{u}-<\bm{u}>)^2>$, $q$ is the flux vector of
``granular heat'', $\bm{\hat{\tau}}: \nabla \bm{u}$ is the tensorial product 
for the granular-temperature generation by shear work 
and $\Gamma$ is the sink term due to dissipation into 
thermodynamic heat (i.e., energy lost in collisions).  
The existing approaches differ
in the constitutive relations that make $\tau$, $q$ and 
$\Gamma$ depend on the other properties $\nabla \bm{u}$, $\rho$, $T$ and on all
the parameters of the system. Apart some heuristic relations (see, for example, ~\cite{Bagnold} or~\cite{Haff}), there are many 
calculations based on the kinetic theory of nonuniform dense gases~\cite{chapman}, making some assumption on the form of the single particle distribution
 function $f(\bm{x}, \bm{v}, t)$, that is the solution of the Boltzmann-Enskog kinetic equation. 
Before~\cite{lun}, and the almost simultaneous~\cite{jenkins}, all the theories had assumed a Maxwellian 
velocity distribution, but the nonzero off diagonal components  in the 
streaming stress tensor (strongly apparent at low solid fractions) indicate
the need of a correction to Maxwellian distribution. In~\cite{lun} and~\cite{jenkins} accurate 
predictions of streaming stresses, in the case of slightly inelastic and 
slightly rough smooth spheres, are obtained. In~\cite{lun} and~\cite{jenkins}
 and in successive studies 
(see \cite{campbell}), a 
certain degree of energy-equipartition breaking is assumed, considering two 
different temperatures for translational and rotational degrees of freedom, 
respectively. 

The increasing power of computers awakened an interest in ``granular
gases'' simulations, that is the investigation of kinetics of granular systems
far from close-packing, e.g. a granular phase diagram has been proposed 
in~\cite{phase} to clarify different regimes and distinguish between a gas-like
phase and two different high-density phases. Results from these simulations 
have been compared to previous granular hydrodynamics predictions, showing
a disagreement in various aspects. 

Many simulations have been performed in the ``cooling'' situation 
(\cite{goldhirsch}, \cite{mcnamara92}, \cite{mcnamara93}, \cite{mcnamara94}, 
\cite{mcnamara96}, \cite{luding98}): the particles evolve with no external forcing, dissipating
in collisions all initial energy. Hydrodynamic predicts a time-scaling of
granular temperature $T \sim t^{-2}$ under the assumption of Maxwellian 
velocities at all times, but strong departures from this law 
are observed when (at fixed volume) the number of particles grows or when 
the restitution coefficient $r$ decreases 
~\cite{mcnamara92}. When $N(1-r)>>1$~\cite{mcnamara93}
 (where $N$ is the number of particles) it is 
found that the clustering instability (this can be derived~\cite{goldhirsch}
 from Jenkins \& Richman hydrodynamics~\cite{jenkins}) may degenerate in 
the so-called {\em inelastic collapse}, as particles may be trapped in a
sequence of infinite collisions in a finite time (i.e. a divergence of 
collision rate). Inelastic collapse is found in one-dimensional and 
two-dimensional~\cite{mcnamara94} simulations.
 Furthermore, equipartition between rotational and translational
energy is found to be broken in cooling kinetics (see~\cite{luding98} and 
references therein). 

Strong equipartition breaking is found in another class of models, that of 
driven granular gases, where the dissipation of energy due to collisions is 
balanced by an external source (in realistic situations one has to vibrate
or shake a granular system, to keep it \emph{alive}). We could divide these
models in two sub-classes: elitary and democratic models, referring to the
quantity of particles receiving energy from the external source. In elitary
(one dimensional~\cite{kadanoff}, \cite{grossman96} and two-dimensional~\cite
{grossman96}) models a wall of the container is the unique energy source, 
therefore there are few particles (just one, in 1d) that transport energy from
the source to the rest of the system. In democratic models (one 
dimensional~\cite{mac}, \cite{swift} and two-dimensional~\cite{peng}) all
particles receive energy, by mean of a Brownian-like random velocity kick 
at every time step. The model we propose in this paper belongs to this 
last sub-class.

Recently the Boltzmann-Enskog equation for granular
kinetic (cooling or driven) has been analyzed, showing that the velocity
distribution is expected not to be Maxwellian. Esipov and P\"{o}schel~\cite
{esipov} have found, for cooling inelastic hard spheres, exponential tails
while van Noije and Ernst~\cite{ernst} have obtained the same tails in the 
cooling regime and $\sim exp(-Av^{3/2})$ tails in the (democratic) 
driven regime.

A non-Maxwellian behavior has been, very recently, observed 
experimentally in a vertically driven granular bed \cite{olafsen}. 
The measured velocity distribution in such experiment is in
very good agreement with the results of our simulations.

To conclude this brief introduction, 
one has to remember that, in modeling granular
gases, the 
oversimplifying criterion is, sometimes, misleading. Brilliantov {\em et al.} 
have shown~\cite{brilliantov} that the universally accepted picture of
fixed restitution coefficient is far from being obvious and that the 
behavior of granular gases (self diffusion, as an example) may 
change drastically if 
this coefficient is taken dependent on the impact-velocity. McNamara and
Luding~\cite{mcnamara98} have recently stressed the importance of keeping into
account rotational degrees of freedom as well translational ones, and that of
the particle-wall dissipation. Furthermore, the relevance of choosing the
boundary energy source and the possibility of considering soft particles instead
of hard grains have been investigated by Geisshirt {\em et al}~\cite{geisshirt}.

In section \ref{modello1} detailed results of simulations of a one-dimensional
 model of driven granular gas  (already presented in \cite{lettera}) are 
reported. In section \ref{modello2} a one-dimensional and two-dimensional 
version (in Boltzmann approximation) of the same model is discussed, 
reporting analogous results. Then, in sections \ref{vittorio} and \ref{angelo},
 some theoretical interpretations are proposed in order to understand 
non-maxwellian behavior and try to relate it to the clusterization 
phenomenon. Section \ref{fine} is devoted to conclusions and open problems.

\section{The inelastic hard-rod one-dimensional model}
\label{modello1}

 Such class of models originates from the seminal 
paper of Du, Li and Kadanoff \cite{kadanoff} 
who considered $N$ identical hard rods 
confined between a thermal and a reflecting wall.
In this model one has a  
statistically steady state as the result of 
the balancing between the dissipation of the kinetic energy 
due to the collisions between the rods and the 
energy reinjection due to the thermal wall; the latter supplies
energy only to the last particle,  
which in turn transfers energy and momentum to the rest of the
system, producing a somehow trivial cluster near the opposite wall. 
Such a state represents a breakdown of the equipartition 
of the energy in a stationary non-equilibrium system; however, its
existence comes about as an artifact, since it is due to the 
peculiarity of the boundary conditions.
In fact, in the model introduced in ref. \cite{kadanoff}
 the mean kinetic energy per particle, 
\beq 
E=\frac{1}{2N} \sum_{i=1}^N <v_i(t)^2> \label{kad_ec},
\end{equation}
and the mean dissipated energy per particle per unit time,
\beq 
W=\frac{1}{\Delta t} \sum_j <(\Delta E)_j> \label{kad_ed},
\end{equation}
(where $(\Delta E)_j$ is the energy loss during the $j_{th}$ collision 
occurred in the time interval  $[t- \Delta t/2,t+ \Delta t/2]$, 
and $< \; >$ is the time average),   
decay exponentially  with the number $N$ of particles,
as shown in figure (\ref{fig_nolimit}).
 Finally, within the Kadanoff model only a small region of parameter 
space can be explored: since $r$, the restitution coefficient, defined 
below, must satisfy the inequality
$N(1-r)<1$ in order
to avoid inelastic collapse \cite{goldhirsch}, \cite{mcnamara93}.

 Williams {\it et al.} \cite{mac} proposed an alternative heating
mechanism. The idea is to supply kinetic energy to
every particle by means of a random acceleration at every 
time step. Since the dissipation due to inelastic collisions 
is not  effective in
balancing the increase of energy coming from the random kicks 
(the latter is independent from the velocities, while the former 
is proportional to them), the authors subtract the average velocity of the 
center of mass of the system from the
velocity of each particle
at every time step in order to avoid
 energy divergence and total non-conservation of the total momentum. 
Though this method is numerically efficient, 
it does not appear realistic from a physical point of view.

Hereafter,
we propose a model \cite{lettera} consisting of $N$ identical particles of 
mass $1$ on a ring of length
$L$. Between collisions, each particle obeys to the following Langevin equation \cite{c43}:
\bea 
\frac{dv_i}{dt}=-\frac{v_i}{\tau}+\sqrt{\frac{2T_F}{\tau}}f_i(t)
 \label{lang1} 
\eea
\bea 
\frac{dx_i}{dt}=v_i(t) \label{lang2} 
\eea
where $1 \le i \le N$, $\tau$ and $T_F$ are the relaxation time due to viscous effects and the thermal bath temperature respectively, $f_i(t)$ is a 
standard white noise with zero average and $<f_i(t)f_j(t')>=
\delta_{ij}\delta(t-t')$.

In addition to these equations, the particles mutually collide
 according to the following rules:

a) only binary collisions are considered, 

b) each collision is instantaneous,

c) the post-collisional velocities are related
to the pre-collisional ones by the equations:

\bea v_i' & = & \frac{1-r}{2}v_i+\frac{1+r}{2}v_j \nonumber \\
    v_j' & = & \frac{1+r}{2}v_i+\frac{1-r}{2}v_j
\eea

where $r$ is the \emph{restitution coefficient}. In this way, 
the momentum is 
conserved in the
collisions, while the kinetic energy of center 
of mass is rescaled by $r^2$, i.e.
\bea (v_i'-v_j')^2=r^2(v_i-v_j)^2 \eea
The elastic case is $r=1$, while for $r=0$ the colliding particles have no 
relative motion after the collision: they move together
with the velocity of the center of mass. It has 
to be noted that, in one dimension, the size of the  particles
is not a significant
parameter, because of the ``hard'' nature of collisions: 
the particles never deform
(this effect is kept into account in the restitution 
coefficient picture) and therefore only the length of spacings between
particles is important, that is $L$ (see \cite{kadanoff})

In absence of collisions, each particle would perform a Brownian motion 
reaching, for $t>>\tau$, a stationary state with a gaussian 
velocity distribution:

\beq 
P(v_i)=\frac{1}{\sqrt{2 \pi T_F}} \exp \left( -\frac{v_i^2}{2T_F} \right)
\end{equation}

and a diffusive behavior:

\beq <(x_i(t)-x_i(0))^2>=2Dt  \end{equation}

where $D=T_F \tau /2$ is the diffusion coefficient. 
The interpretation of the model is straightforward: the environment 
supplies kinetic
energy to the system as a thermal bath at temperature $T_F$. 
The viscous term 
(with characteristic time $\tau$) is naturally introduced 
to take into account
different friction effects, such as contact friction with boundaries, 
particle-fluid interaction, tangential inter-particle friction and energy 
transfer among different degrees of freedom. Experimental fluidized beds, 
see for example \cite{fluidized}, are an example of phenomena in which 
a viscous damping and a noisy term are naturally present. 
As noted before, in \cite{olafsen}
is presented an experiment showing strong analogies with our model.

When collisions are considered, another characteristic time 
emerges, that is 
the average collision time $\tau_c$ between two successive 
encounters. An estimate of $\tau_c$, as a function of average density 
and typical velocity, is 
 
\beq 
\tau_c \sim \frac{L}{2N \sqrt{<v^2>}} \label{stima_tauc}. 
\end{equation} 

It is natural to assume that $<v^2>$ reaches a stationary value with 
statistical fluctuations (of order $\sim 1/N$), as it is observed in 
simulations. In the following we shall refer to the quantity

\beq 
T_g=<v^2>=\lim_{(T-T_0) \rightarrow \infty}
\frac{1}{(T-T_0)N}\sum_{i=1}^N \int_{T_0}^T v_i(t)^2 dt 
\end{equation}

as to the \emph{granular temperature} of the system.
 Note that the system is not at equilibrium, 
therefore $T_g$ is not a temperature in a proper thermodynamic sense: it may be different, if one measures it at different scales or in different subsystems 
(as it will be shown later).

In all the simulations performed, we use $L/N=1$ and $T_F=1$ and the measured
$T_g \leq T_F$ is almost always 
found greater than $0.1$ (and never drops below $0.03$). 
>From eq. (\ref{stima_tauc}) we can estimate $0.5< \tau_c <5$.

The presence of two time scales ($\tau$ and $\tau_c$) 
leads to two different regimes. 
As $\tau_c$ varies in a small range (less than one order of magnitude), 
we could tune the parameter $\tau$ to observe these two phases:

A) When $\tau<<\tau_c$ the effect of collisions is rapidly 
overwhelmed by the
Brownian motion (i.e. collisions are rare events) and the system
behaves as a collection of weakly interacting random walkers or, equivalently,
as an ideal gas in equilibrium at a temperature $T_g$ not far from 
the temperature 
$T_F$ of thermal bath (one can be convinced of this also looking to the Boltzmann equation of the system, reported in the next section).

B) In the opposite limit $\tau_c<<\tau$ 
the collisions dominate the dynamics and
strongly compete against the driving mechanism. In this regime a statistically
stationary state is still observed, in the sense that macroscopic averages are
well defined, and interesting phenomena emerge: 

a) a strongly inhomogeneous spatial distribution 
(\emph{clusterization}) 

b) a deviation of velocity 
distribution from gaussian behavior. These phenomena are
more and more pronounced with decreasing values of the restitution coefficient
$r$.

The simulations have been performed using a fixed step $\Delta t$ integration of 
eqs. (\ref{lang1}) and (\ref{lang2}) where $\Delta t << \tau_c$ and an event
driven check of collisions during every time step. For low values of $r$ 
we observed an exponential decreasing of collision time, 
much shorter than
the integration time $\Delta t$. We discarded these simulations, interpreting
them as examples of inelastic collapse~\cite{mcnamara93}. 
The critical value of $r$, for the appearance of collapse, increases with
$\tau$: in the limit $\tau \rightarrow \infty$ the thermal bath disappears
and the system becomes a granular cooling model with critical value, for 
inelastic collapse, estimated by $N(1-r) \sim 1$ as noted before.

In figure (\ref{fig_ewriepilog1d}) we report 
  $T_g$ and $W$ vs. the restitution coefficient $r$ for different $\tau$.

A simple relation between $T_g$ and $W$ may be obtained. The variation of the kinetic energy due to Langevin dynamics is
\begin{eqnarray}
 \label{deltaTg} (\delta E(t))_{Lang} & = &
\frac{1}{2N} \sum_{i=1}^N(v_i(t)+\delta v_i(t))^2-
\frac{1}{2N}\sum_{i=1}^Nv_i^2(t)= \\ 
\nonumber && = \frac{1}{2N}\sum_{i=1}^N(\delta v_i(t))^2+ \frac{1}{2N}\sum_{i=1}^Nv_i(t) \delta v_i(t)  
\end{eqnarray}
where $\delta v_i$ is the velocity variation 
during a time interval $dt$ in equation (\ref{lang1}), from which we
obtain the relations:
\beq \label{deltav^2} \lim_{dt \rightarrow 0} \left \langle \frac{\delta v_i(t))^2}{dt} \right \rangle =
      \sqrt{\frac{2T_F}{\tau}}
\end{equation} 
\beq \label{deltav} \lim_{dt \rightarrow 0} \left \langle \frac{v_i(t) \delta v_i(t)}{dt} \right \rangle=
     -  \frac{\langle(v_i(t))^2 \rangle}{\tau} \end{equation}
where the $<...>$ average is taken over different realizations of stochastic 
process $f_i(t)$. Recalling the definition of $T_g$, 
 using equations (\ref{kad_ec}),
(\ref{kad_ed}), 
and inserting eqs. (\ref{deltav^2}) and (\ref{deltav}) into
eq. (\ref{deltaTg}), and 
assuming the ergodicity, one obtains:
\beq W=\frac{T_F-T_g}{\tau} \label{relazionewt}  \end{equation}
The numerical check of such relation is shown in figure
 (\ref{fig_relazionewt})

Though the system is statistically stationary, the 
instantaneous density of particles is rapidly evolving. To get an idea 
of different density profiles in 
the two regimes (homogeneous and clusterized), 
look at fig.(\ref{fig_density1d}).

The density distribution becomes fractal in the clusterized regime, as can be
verified measuring the correlation dimension $d_2$ \cite{gra_pro} that we
computed from the correlation function
\beq C(R)=\frac{1}{N^2 (T-t_0)}\int_{t_0}^T dt \sum_{i,j} \theta(R-|x_i(t)-x_j(t)|) \sim R^{d_2} \label{co_fun} \end{equation}
with $t_0$ the time after which one can assume the system is in a typical situation and $T$ is the duration
of the simulation. C(R) is shown for homogeneous and clusterized situations in
figure (\ref{fig_gp1d}). In figure (\ref{fig_gp1dtot})
 is presented a summary of
$d_2$ measurements as function of parameters $\tau$ and $r$. 

The clusterization may quantitatively characterized by means of an 
\emph{entropy}
defined as
\beq h_M=-\sum_{j=1}^M \frac{m_j}{N} \log \frac{m_j}{N} \end{equation}
where the ring of length $L$ is divided in $M$ equal \emph{boxes} (i.e., segments) 
and $m_j$ is the number of particles in the $j_{th}$ box. The 
entropy $h_M$ attains its maximum value $h_M=\log M$ when $m_j=N/M$ for every 
box $j$. $h_M$ decreases as the density distributions becomes more and more 
clusterized. For a non-clusterized (but fluctuating) density, we have
a Poisson distribution for $m_j$, that is (with $\lambda=N/M$)
\beq f(m_j)=\frac{\lambda^{m_j}}{m_j!}\exp(-\lambda) 
\label{poissonh} \end{equation}
from which it can be (numerically) calculated the effective entropy for 
homogeneous regime $h_M^*$. In figure (\ref{fig_entropy1d}) are presented 
many measurements of $H_M/H_M^*$ where $H_M=\exp(<h_M>)$, $H_M^*=\exp(<h_M^*>)$ 
and $< \; >$ is the time average. The quantity $H_M/H_M^*$ basically gives
an indication of the fraction of non-empty boxes in a typical snapshot.

In figure (\ref{fig_distvel1d}) is shown the distribution of velocities, 
obtained sampling the velocities of all particles 
for very long times, in the two different regimes 
(a quasi-equilibrium case with $\tau=0.01$, 
$r=0.99$ and an out of equilibrium case 
with $\tau=100$, $r=0.7$). In the quasi-equilibrium
regime the distribution is very well fitted by a gaussian. 
As a general result, when $\tau>>\tau_c$ 
the velocity distribution ceases to be gaussian and the
high velocity tails decay more slowly to zero. The deviation becomes more
pronounced as the restitution coefficient $r$ decreases. In the figure a 
theoretical fit is also plotted for the non-gaussian distribution. This fit
will be discussed below, in section \ref{vittorio}.

In figure (\ref{fig_denfluct1d}) the density distributions in the 
homogeneous and clusterized regimes 
are shown (respectively $\tau=0.01$, $r=0.99$ and
$\tau=100$, $r=0.7$, with $N=300$), i.e. $f_M(m)$ where $m$ is the number
of particles in a box when the ring is divided in $M$ boxes (this
distribution, as all the others, is obtained sampling data for very long times). The homogeneous regime is very well fitted by a Poisson distribution, as
noted in equation (\ref{poissonh}). The clusterized regime presents an exponential
long tail and a power law for the low density boxes: the function
$\frac{1}{m}e^{-cm}$ with $c=0.14$ fits very well the data and is
consistent with the theoretical interpretation given in
 section \ref{vittorio}.

The figure (\ref{fig_enfluct1d}) represents the \emph{
box granular temperature} $T_M(m)$ as a function of the number of particles $m$ in a box:
\beq T_M(m)=\frac{1}{m}\sum_{j=1}^m (v_j-<v>_m)^2 \end{equation}
where $<v>_m$ is the average velocity (typically close to zero when $m>>1$) in the box and $M$ is the number of boxes on the ring of length $L$. The figure shows this
function (averaged over very long times) for both the
regimes: in the gaussian case (with $\tau=0.01$ and $r=0.99$) we observe that
$T_M(m)$ is a constant, while in the non-gaussian case ($\tau=100$, $r=0.7$)
it is a power law, i.e. $T_M(m) \sim m^{-\beta}$ 
with $\beta=0.5$. The exponent $\beta$ 
depends on the values of 
$\tau$ and $r$.

\section{The Boltzmann Equation approximation}
\label{modello2}

A natural question now arises: can we expect that the above results 
are general and independent from the dimensionality 
of the system? Or are these an artifact of the one
dimensional dynamics? An answer may come
from the Boltzmann equation for the one particle distribution $P(x,v,t)$ 
\cite{k81}:
\beq \pdev{P}{t}+\pdev{(vP)}{x}-\frac{1}{\tau}\pdev{(vP)}{v}-\frac{T_F}{\tau}
\pdev{^2P}{v^2}=\left. \pdev{P}{t} \right|_{coll} \label{boltzmann}
\end{equation}
\begin{eqnarray}
 \left. \pdev{P}{t} \right|_{coll}=\frac{4 \Lambda}{(1+r)^2} 
     \int dv' |v'-v| P(x,v',t) P(x,(2v-(1-r)v')/(1+r),t)
     - \nonumber \\ 
- \Lambda \int dv' P(x,v',t)P(x,v,t)|v'-v| \label{col_int}
\end{eqnarray} 
where $\Lambda \sim 1/\tau_c$ is the mean collision rate per particle. In the
limit of elastic collisions ($r=1$) the collision integral (\ref{col_int})
disappears and the stationary solution of eq. (\ref{boltzmann}) is
 $P(x,v,t) \propto \exp(-v^2/2T_F)$.
This is related to the fact that in the elastic limit a collision between 
two particles is nothing but a change of the labels of the two particles
and therefore the collisions are not relevant at all.

The main approximation in eq.(\ref{boltzmann}) is the Boltzmann 
\emph{Stosszahlansatz}, according to which the correlation between two
close particles is neglected and one writes:
\beq 
P_2(x,x',v,v',t)=P(x,v,t)P(x',v',t) 
\end{equation}

As equation (\ref{boltzmann}), as far as we know, cannot be solved
 analytically,
we consider a stochastic process
based on the Bird algorithm~\cite{Bird},
whose statistical features are identical to those of the Boltzmann
equation. $N$ particles move on a torus 
(in $d$ dimensions) of linear size $L$ (i.e., the area of the d-torus
 is $L^d$).
 The time is discretized in
intervals of duration $\Delta t$. A collision 
time $\tau_c$ is fixed \emph{a priori}: 
this means that two particles collide, during $\Delta t$, with a probability 
$p=\Delta t/\tau_c$ ($\Delta t << \tau_c$, as usual). 

At each discrete time $t_k=k\Delta t$, positions and velocities are
upgraded according to eqs. (\ref{lang1}) and (\ref{lang2}).
Then for each particle $i$ a random number $y$ is extracted out of a uniform
distribution in the interval $[0,1]$: if $y>p$  no collision occurs, 
otherwise the particle $i$ collides with a particle $j$ such that 
$|x_i(t_k)-x_j(t_k)|<l$ ($l<<L$), chosen with probability proportional to 
$|v_i(t_k)-v_j(t_k)|$. The collision rule, in one dimension, is the same as
before and it is extended to the two and three dimensional cases in a natural
way: after a collision $\mathbf{v_i}'$$-\mathbf{v_j}'$$=r\hat
{\epsilon}$$(\mathbf{v_i}$$-\mathbf{v_j})$ where $\hat{\epsilon}$
 is a unit vector with random orientation.

It can be demonstrated~\cite{wag} that for this process,
in the limit $N \rightarrow \infty$, $p \rightarrow 0$, $l \rightarrow 0$,
$\Delta t \rightarrow 0$, the evolution of the probability distribution
  $P(x,v,t)$ of the Bird process is governed by the
 Boltzmann equation (\ref{boltzmann}).

In this model there are two parameters, $\tau_c$ and $l$. The first one was
already an observable of the previous model (where it was almost constant). The
second one, instead, represents the collision range and replaces the
radius of the particles, so it is related, in some way, to the total volume 
fraction $\nu=kN/L^d$ with $k$ the volume of one particle, which is not 
explicitly considered in this model and in the previous one. Furthermore, we
cannot expect to find a fractal scaling at a range lower than $l$, and, 
moreover, no inelastic collapse can be observed, as short range correlations 
are neglected. It should be noted, finally, that the imposed collision
time $\tau_c$ is larger (but of the same order of magnitude) than the one
really measured
in the simulations, $\tau_c^*$, because
a particle collides with probability $p=\Delta t/\tau_c$  only if there is
another particle at distance lower than $l$. One can expect that  
$\tau_c^* \rightarrow \tau_c$ as the clusterization becomes more and more
pronounced.

In all the simulations of this model we kept $T_F=1$ and $N/L^d=1$, we rarely
changed $\tau_c$ and $l$ and we explored the space of parameters $\tau$ and $r$
as in the previous section. All the results previously obtained were 
recovered in this approximation with $1\leq d \leq 3$, showing that they are 
general properties of a granular system subject to such a heating mechanism 
as that of eqs. (\ref{lang1}), (\ref{lang2}). A rapid overview of significant 
measurements, in one and two dimensions, follows.

We stress the fact that also in this model the system reaches a statistically
stationary state after a transient, and also in this model there are two 
different regimes: the quasi-equilibrium regime ($\tau<<\tau_c$) and the
out-of-equilibrium one ($\tau>>\tau_c$). 

The clusterization phenomenon is shown in figures (\ref{fig_de2}), 
(\ref{fig_fr1}) and (\ref{fig_fr2}) where the density snapshots 
and the correlation functions $C(R)$ (defined 
in eq. (\ref{co_fun})) 
are presented. It is observed the predicted reduction of the
fractal scale (more evident in the two-dimensional model) due to Boltzmann
approximation.

The existence of a good thermodynamic limit is shown in the figures
 (\ref{fig_lt1}) and (\ref{fig_lt2}) where the $N$-dependence of some 
observables is plotted:
we show the granular temperature $T_g$, the fractal dimension $d_2$
and the collision rate $\chi=1/\tau_c^*$, all in the out-of-equilibrium 
regime.

The same quantities are plotted in figures (\ref{fig_riep1}) and (\ref{fig_riep2})
 against the restitution coefficient $r$ in the
same regime (these plots are analogous to those of figures (\ref
{fig_ewriepilog1d}) and (\ref{fig_gp1dtot})). Note that $\tau_c^*$ approaches
$\tau_c$ when $r$ decreases, as it is expected. The diffusion coefficients,
also plotted in those figures, will be discussed in the conclusion.

The distributions of velocities are presented in figure (\ref{fig_ve1}) and 
(\ref{fig_ve2}). The non-gaussian behavior in the second regime ($\tau>>\tau_c$)
is still clearly observed. In figure (\ref{fig_ve_de}) a distribution of 
velocities restricted in the following way is presented: we
sampled the velocities of one particle only when there were other $m$ ($m=1$
and $m=5$) 
particles in a box of radius $R$ ($l<<R<<L$) centered on that particle. 
This is a sort of velocity distribution at fixed density. The plot shows a
less pronounced deviation from the gaussian, in agreement with the statement
(discussed in section \ref{vittorio}) that there is a local equilibrium with a temperature that
depends upon the local density, in order to have a stationary distribution of
clusters.

The analogue of the figure (\ref{fig_denfluct1d}) and (\ref{fig_enfluct1d}) are
the figures (\ref{fig_df1}), (\ref{fig_df2}), (\ref{fig_ef1}) and (\ref{fig_ef2}).
Again the density distribution $f_M(m)$ is a Poisson function when the system
is not clusterized and a function of the form $exp(-\alpha m)/m^{\beta}$ in
presence of clustering. The \emph{box granular temperature} $T_M(m)$ 
still presents
a constant behavior (as expected) at equilibrium and a power law $m^{-\gamma}$
in the non-gaussian regime. It does not seem possible to easily 
find a relation between $\alpha$, $\beta$, $\gamma$ and the other 
observables (as $d_2$ or $T_g$).

In summary, the exact model is perfectly reproduced in this Boltzmann 
approximation, at large $N$, not only confirming the existence
of a thermodynamic limit, but also showing that this system may be 
studied analytically in order to better understand this kind of driven 
granular kinetics. In the following sections some theoretical interpretations
 will be discussed.

\section{A model to explain clusterization}
\label{vittorio}
In this section we address the problem of the microscopic origin of the 
clusterization. In order to do that, we study a class of models 
in which the system is composed by $M$ boxes and $N$ particles
in a mean-field model, assuming that the boxes have infinite connectivity.
One starts with a certain configuration and let the system evolve
with an exchange dynamics in which, at each time step,
one  particle moves from one box to another, both boxes being chosen
randomly. The probability for each single exchange is model-dependent and 
it will be our tuning-parameter to scan the different phenomenologies.
Our goal is to understand in a quantitative way how the microscopic 
dynamics affects the clustering properties of the system.
In particular we shall try to recover the results, obtained in the framework 
of the models previously introduced, for the density distributions in 
the clusterized and homogeneous cases (see Figs. (\ref{fig_denfluct1d}),
(\ref{fig_df1}). (\ref{fig_df2})).

The models are defined in terms of master equations for the
probability $P_m$ of having a box with $m$ particles, assigning
 transition rates 
for landing in a box with $m$ particles $W_{in} (m)$ and for leaving a box
with $m$ particles $W_{out} (m)$. It must be 
\begin{equation}
W_{in}(N)=W_{out}(0)=0
\end{equation}
and the normalization conditions must be satisfied:
\begin{equation}
\sum_{m=0}^{\infty} P_{m} =1 \hspace{2.0cm}
\sum_{m=0}^{\infty} m P_{m} =\frac{N}{M}= \lambda \hspace{2.0cm}
\sum_{m=0}^{\infty} W_{in} (m) M P_m  = 1
\label{norm}
\end{equation}

The general question is:
 what is the asymptotic stationary distribution for the 
average number of boxes with $m$ particles, $P(m)$?

The simplest case we can consider is the one in which each single movement
is independent of the state of the departing and of the landing box.
In this case there is no bias in the movements and 
$W_{in} (m)$ and $W_{out} (m)$ do not depend upon $m$:
\begin{equation}
W_{in} (m) = W_{out} (m)= \frac{1}{M}
\end{equation}
and the general master equation reads
\bea
\label{master}
M^2 \frac {dP_{m}}{dt}& = & P_{m-1}(P_{m-1}-\frac{1}{M})+ 
 2 P_{m+1} P_{m-1}+P_{m+1} (P_{m+1}-\frac{1}{M})+
\\ \nonumber && +P_{m+1} (1-P_{m}-P_{m+1}-P_{m-1})+ 
 (1-P_{m}-P_{m-1}-P_{m+1}-P_{0}) P_{m-1} +
\\ \nonumber  && -P_{m-1}P_{m}-2P_{m}(P_{m}-\frac{1}{M})-
P_{m}P_{m+1}- P_{m}(1-P_{m-1}-P_{m}-P_{m+1})+
\\ \nonumber && - (1-P_{m}-P_{m+1}-P_{m-1}-P_{0})P_{m} 
\hspace{3.0cm} \mbox{for  $0 < m < N$}
\\ \nonumber
M^2 \frac{dP_{0}}{dt}& = & P_{1}(1-P_{0}-\frac{1}{M})
-  (1-P_{1}-P_{0})P_{0}\\
M^2 \frac{dP_{N}}{dt}& = & P_1P_{N-1}-P_{N}(1-\frac{1}{M})
\end{eqnarray}

In the limit of $M >> 1 $ one can neglect the $\frac{1}{M}$ terms in
the right hand side of eq. (\ref{master})
and easily get the stationary solution ($\frac{dP_m}{dt}=0$)
\begin{equation}
P_{m} = A e^{-c m}
\label{expo}
\end{equation}
with $A = 1- e^{-c}$ corresponding to the normalization
condition $ \sum_0^{\infty} P_{m} =1$ and where $c$ is a constant 
depending on $N$ and $M$: $c = ln(1+\frac{1}{\lambda})$ with
$\lambda= \frac{N}{M}$.

This result has to be compared with the probability $f_M(m)$
in the non-clusterized case  of the previous sections.
In order to do this it is necessary to recall that this result 
has been obtained with a small value of the number of boxes $M$.
This means that one is very far from the limit $M >> 1$ and 
this situation corresponds to a sort of coarse graining in the 
system in which each box (big box) is actually composed by 
a certain number of small boxes (whose number is such that $M>> 1$).
The problem can thus be formulated in the following way:
given a system of $N$ particles distributed in $M_{small}$ boxes with 
the distribution $P_{m}$ given by eq. (\ref{expo}), what is the distribution 
$P^*_{m}$ for the particles in a system of $M_{big}$ boxes each one composed 
by $R$ ($R=M_{small}/M_{big}$) small boxes? 
The resulting distribution is easily written as
\begin{equation}
P^*_{m}= \sum^* 
\prod_{i=1}^R
P_{m_i}= A^R e^{-c m} F(m,R),
\label{poisson}
\end{equation}
where $\sum^*$ indicates the sum on the $\left\{ m_1,...,m_R \right\}$
such that $\sum_{i=1}^R m_i = m$,  
$F(m,R)$ is the number of ways of distributing $m$ 
particles in $R$  boxes and it is given by \cite{cassi}:
\begin{equation}
F(m,R)= \left(
\begin{array}{c}
{m+R-1} \\
{m}
\label{comb}
\end{array}
\right)
\end{equation}
With the help of (\ref{comb}) and using the Stirling formula,
 the expression (\ref{poisson}) becomes (for $R>>N>>1$)
\begin{eqnarray}
P_m^* & = & A^R e^{-cm} \frac{(m+R-1)!}{m!(R-1)!} \approx 
 A^R e^{-cm} \frac{(m+R-1)^m
       (m+R-1)^{R-1}}{m!(R-1)^{R-1}} \approx \\
\nonumber && \approx A^R e^{-cm} \frac{R^m}{m!} = 
\frac{A^R}{m!} e^{-m (\ln(1+\frac{M_{small}}{N})-\ln
  \frac{M_{small}}{N} - \ln \frac{N}{M_{big}})} \approx \\
\nonumber && \approx e^{-\lambda^*} \frac{(\lambda^*)^m}{m!} 
\end{eqnarray}
It has been used the definition of $R$, the fact that
$c=\ln(1+M_{small}/N)$ and that $M_{small}/N>>1$. In the last passage
$\lambda^*=N/M_{big}$ has been introduced and $A^R$ has become
$e^{-\lambda^*}$, as can be verified when $\lambda^{-1}=M_{small}/N>>1$.
It has been shown, therefore, that the coarse grained version of the
solution  of (\ref{master}) is exactly the Poisson distribution found
in the simulations, in the non-clusterized regime
 (see figs. (\ref{fig_denfluct1d}), (\ref{fig_df1}) and (\ref{fig_df2})).

Let us consider now one case where the transition rates for the 
particle jumps depend on the contents of the departing and landing boxes.
This corresponds to impose some sort of bias to the system that could well 
reproduce the situation one has in the clusterized cases due to the 
inelasticity.
We consider in particular the following case, defined by the transition rates:
\begin{eqnarray}
\label{biased}
W_{in} (0)& =& \frac{1}{M} \nonumber \\  
W_{in} (m)& =& (1-P_{0}) \frac{m}{N} \hspace{2.0cm} 
\mbox{for $0 < m < N$}\\
W_{out} (m)&=& \frac{m}{N} \hspace{3.3cm} \mbox{for $0 < m \le N$} \nonumber
\end{eqnarray}

These transition rates, that satisfy the relations (\ref{norm}), 
have the following interpretation. The probability to 
land on a box containing already $m$ particles is proportional to the 
number of particles because this mimics the inelastic collision 
with a cluster of $m$ particles. On the other hand the departure from a box 
containing already $m$ particles has a probability proportional to $m$ 
because the probability to select one particle in that particular box is 
proportional to $m$.

Neglecting as usual the terms of the order of $\frac{1}{M}$, and after
simplifications, the stationary master equations write:
\begin{eqnarray}
P_{m+1} (m+1) + (1-P_{0}) (m-1) P_{m-1}- P_{0} m P_{m} & = & 0 
\\ \nonumber P_{1} - \frac{N}{M} P_{0}\hspace{0.0cm}  & = & 0
\\ \nonumber  \frac{P_1}{N} P_{N-1} (1-P_{0}) \frac{N-1}{N} -
      P_N(1-\frac{1}{M})\frac{1}{M} \hspace{0.0cm} & = & 0
\end{eqnarray}
The solution in this case is given by:
\begin{equation}
P_{m} = A \frac{1}{m}  e^{-\alpha m} 
\hspace{1.0cm} P_{0} = 1 - e^{-\alpha}
\label{expo2}
\end{equation}
with $A = \lambda (e^{\alpha}-1)$ and $\lambda= \frac{N}{M}$.
$A$ and $\alpha$ are related by an implicit equation obtained imposing 
the condition $\sum_0^N P_{m}  = 1$, that in the limit $N \rightarrow \infty$
becomes
\begin{equation}
1-e^{-\alpha}-A*\ln(1-e^{-\alpha})=1
\end{equation}
 In the clusterized case we expect
the solution to be self-similar, in the sense that $P_m$ has the same
behavior of $P_m^*$, and the coarse graining previously performed
should not change the solution (\ref{expo2}), apart a rescaling of
$\lambda$ and $\alpha$. 

It must be noted that, as $A$ must be
finite, when $N \rightarrow \infty$ (and $M$ is fixed) $\alpha$ has to
go to zero, while $\alpha$ diverges when $N/M$ goes to zero. It is
natural to think to $\alpha$ as to the inverse of the characteristic
 ``mass'' of
a cluster, that is the typical number of particles in it.
In this sense
the term $exp(-\alpha m)$ acts as a finite-size cut-off for the
self-similar distribution $P_m \sim 1/m$.

The solution (\ref{expo2}) is in excellent agreement with the numerical results
obtained in the previous sections. In particular 
in the case $N=300$ $M=100$ of the one-dimensional model of Sect.II
one recovers the density distribution with the correct value of 
$\alpha \simeq 0.14$ (see fig. (\ref{fig_denfluct1d})). 

To get the other observed  behaviors of density
 distribution $P_m \sim e^{-\alpha m}/m^{\beta}$ (see 
figs. (\ref{fig_df1}) and (\ref{fig_df2})), it is enough to change the
transition rates appearing in eqs. (\ref{biased}) into the following:
\begin{eqnarray}
W_{in} (0)& =& \frac{1}{M}\\
W_{in} (m)& =& \mu (1-P_{0}) m^{\beta} \hspace{2.0cm} 
\mbox{for $0 < m < N$}\\
W_{out} (m)&=& \mu m^{\beta} \hspace{3.3cm} \mbox{for $0 < m \le N$}
\end{eqnarray}
where $\mu$ is a normalizing constant:
\begin{equation}
\mu=\left( M \sum_i^N P_m m^{\beta} \right)^{-1}.
\end{equation}

Now, we can go a step further relating the clustering
properties of the system to
the velocity distribution. In order to do that we consider the
following quantities: the distribution of boxes, $f_M(m)$,
containing a given
number $m$ of particles  and the velocity variance $T_M(m)$, 
in a box occupied by $m$ particles. 
We consider first the non-clusterized case 
($\tau << \tau_c$ and $r \simeq 1$). Within this regime we find from the 
simulations that:

\begin{eqnarray}
T_m^{elas}(m)& \simeq & const.\\
f_M^{elas} (m)& = & \frac{\lambda^{m} e^{- m} }{ m! }.  
\label{gauss}  
\end{eqnarray}
By assuming in each box a gaussian
velocity distribution with a constant variance $T_M^{elas}(m)$
it turns out that the global velocity distribution $P^{elas} (v)$
is gaussian.  Let us recall that the Poisson
distribution is the one associated with a process  
of putting independently $\lambda N$ particles into $N$ boxes.

Let us turn to the non-elastic case. If 
$\tau=100$ and $r=0.7$, considering the occupied 
boxes ($m > 0$),
we obtain from the simulations the following relations 
\begin{eqnarray}
T_M^{inel}(m)& \sim & m^{-\beta} \label{inel1}\\
f_M^{inel}(m)& = & \frac{e^{- \alpha m} }{ m },  
\label{inel}  
\end{eqnarray}
with $\beta\simeq 0.5$ and $\alpha\simeq 0.14$.
Let us compute from these scalings the global velocity distribution. 
Taking into account that the spatial probability distribution of the particles 
is $f_M(m)$ and assuming that their local
velocity distribution is gaussian, but
with a variance $T_M(m) \simeq  
m^{-\beta}$ which depends on the occupancy,
we obtain, for the global velocity distribution $P_{inel}(v)$,
and in the continuum limit:
 \begin{equation}
\label{numerica}
P_{inel}(v) \simeq \sum_{m=1}^{\infty} e^{(- \frac{v^2 m^{\beta}}{2})} 
e^{-\alpha m}.
\end{equation}
We stress how the the
distributions measured in the simulations are in
very good agreement (see the dashed line in fig. (\ref{fig_distvel1d}))
 with the
 numerical computation of eq. ({\ref{numerica}), which, in summary, has been
   obtained under only the following hypothesis:   
\begin{itemize}
\item[({\bf{i}})] non-Poissonian distribution for the box occupancy 
$f_M(m) \propto e^{-\alpha m}/m$; 
\item[({\bf{ii}})] gaussian distribution of velocities in each box with
a density-dependent variance $T_M(m) \propto m^{-\beta}$.
\end{itemize}

The hypothesis about the scaling relation between the velocity
variance (i.e. $T_M(m)$) and the local density, apart from being justified numerically,
can be understood in the following way.
The stationarity and the scale-invariance of the cluster distribution,
implies a certain distribution of lifetimes for the clusters. In particular 
each cluster has a lifetime which is inversely proportional
to its size. The scale-invariant cluster-size distribution
thus implies a scale-invariant distribution for the lifetimes.
The cluster lifetime is strictly related to the variance
of the velocity distribution inside the cluster itself. In order to ensure the stability of a
cluster in a stationary state we have to require that the velocities 
of the particles belonging to it are not too different, or equivalently
that the variance of the distribution is smaller the higher 
the density. So, given a 
scale-invariant distribution of clusters one would expect a scale-invariant distribution 
of variances, that is $T_M(m) \sim m^{-\beta}$.

In the next section the non-gaussian distribution of velocity will be related to clusterization with the help of a mean field model of driven granular gas.

\section{A model for the clustering and the non-gaussian behavior}
\label{angelo}

In order to shed some light on the relationship between the
spatial clusterization and the anomalous velocity distribution observed
above, we present a simple theoretical model. For sake of notation simplicity, we discuss only the 1d case.
 Let us treat the collisions in a mean-field like fashion
and modify the Langevin dynamics plus collision rules by
the following set of coupled equations for the velocities:
\begin{equation}
\frac{dv_i}{dt}=-\frac{v_i}{\tau}+
\frac{1}{N}\sum_{j=1}^N g(v_i-v_j)
+\sqrt{\frac{2T_F}{\tau}}f_i(t)
\label{angelo1} 
\end{equation}
where the second term in the r.h.s.
 determines the velocity change of the particle 
$i$ due to the collisions with the remaining particles and is chosen
to mimick the inelastic behavior. This requirement poses some constraints
about the form of the function $g(v-v')$:
\begin{itemize}
\item The momentum conservation dictates the
antisymmetric property, $g(v-v')=-g(v'-v)$
\item The inelasticity of the collision process requires 
$g(v-v')(v-v')\leq 0$
\end{itemize}

The Fokker-Planck equation of (\ref{angelo1}) is 
\begin{eqnarray}
\partial_t P_N(v_1,...,v_N,t)-\frac{1}{\tau}\sum_{i=1}^N \frac{\partial}
{\partial v_i} (v_i P_N(v_i,...,v_N,t)-\frac{T_F}{\tau}\sum_{i=1}^N \frac
{\partial^2}{\partial v_i^2} P_N(v_1,...,v_N,t)+ \nonumber \\
+\sum_{i=1}^N \frac{\partial}{\partial v_i} \left[ \frac{1}{N} \sum_{j=1}^N 
g(v_i-v_j) P_N(v_1,...,v_N,t) \right]=0
\end{eqnarray}

>From the above equation, using the fact that in the limit $N \rightarrow 
\infty$ the mean field approximation holds, one 
can obtain an evolution equation for the 1-body velocity probability 
distribution which reads:

\beq 
\pdev{P(v,t)}{t}-\frac{1}{\tau}\pdev{(vP(v,t))}{v}-\frac{T_F}{\tau}
\pdev{^2P(v,t)}{v^2}+\pdev{}{v}\int dv'P(v,t)P(v',t)g(v-v')=0 
\label{angelo2}
\end{equation}

i.e. a sort of self consistent Boltzmann equation.
>From eq. (\ref{angelo2}) one observes that the quantity:

\beq
\int dv'P(v',t)g(v-v')=G(v)=-\pdev{U(v)}{v} 
\label{angelo3}
\end{equation}

which is a function of $v$ and a functional of $P(v)$, can be considered
as an effective force acting on the particle generated by an effective 
potential $U$. Integrating once with respect to the velocity 
the stationary version of 
eq. (\ref{angelo2}) one can obtain the following equation:

\beq
\left(-\frac{v}{\tau}+\frac{T_F}{\tau}\frac{\partial}{\partial v}+ G(v)
\right)P(v)=0
\label{angelo4}
\end{equation}

The solution of eq. (\ref{angelo4}) is:
\beq
P(v) \propto \exp \left(-\frac{\tau}{T_F}\left(\frac{v^2}{2 \tau}+ U(v)\right)
\right)
\label{angelo5}
\end{equation}

In order to make some progress we consider 
the qualitative shape of $g(v_i-v_j)$. In eq. (\ref{angelo1}) the effect of 
collisions between the particles $i$ and $j$ in the unit of time is given by:
\begin{equation}
\frac{d}{dt}(v_i-v_j)|_{coll}=\frac{2}{N}g(v_i-v_j)
\label{angelo6}
\end{equation}
The variation of momentum in an interval $dt$ can be rewritten as 
\begin{equation}
\delta(v_i-v_j)|_{coll}=\delta q_{ij} \cdot (v_i-v_j)
\label{angelo7}
\end{equation}
where $\delta q_{ij}$ is the analogue of $q=1-r$ in the model discussed in
sections \ref{modello1} and \ref{modello2}. The important difference is
 that here $\delta q_{ij}$ 
 represents  the effect of all the collisions during $dt$, and thus can be
 associated to an effective restitution coefficient. 
Eq. (\ref{angelo7}) may be rewritten as:

\begin{equation}
\frac{d}{dt}(v_i-v_j)|_{coll}=\chi_{ij} q \cdot (v_i-v_j)
\label{angelo8}
\end{equation}
where $\chi_{ij}$ is the number of collisions between the $i-th$
and the $j-th$ particles in the unit of time. 
Upon comparing eqns. (\ref{angelo6}) and (\ref{angelo8}) one obtains an
 expression for $g(v_i-v_j)$:
\begin{equation}
\label{gdichi}
g(v_i-v_j)=\frac{2 \chi_{ij} q}{N}(v_i-v_j)
\end{equation}
Now it is easy to understand that
$\chi_{ij}$ is a decreasing function of $|v_i-v_j|$: indeed, a great number of
collisions occurs when the pair $i,j$ belongs to a cluster (where $|v_i-v_j|$
is small), whereas the two particles rarely collide when they are out of a 
cluster (and $|v_i-v_j|$ is high). We can, therefore, make a rough estimate 
of $g(v_i-v_j)$, that is:
\begin{eqnarray}
\label{stima}
|g(v_i-v_j)| \sim \frac{|v_i-v_j|}{\tau_c} \hspace{2cm} \mbox{inside clusters} \nonumber \\
|g(v_i-v_j)| \sim |v_i-v_j|^{\beta '} \hspace{2cm} \mbox{outside clusters}
\end{eqnarray}
where $\beta'<1$.
>From eq. (\ref{angelo3}) it appears that $G(v) \sim g(v-v')|_{v'=0}$ as the
integration has to be performed with respect to the measure $P(v',t)dv'$ 
that is strongly peaked at $v'=0$. Finally, one can conclude from the same
eq. ({\ref{angelo3}) that
\begin{eqnarray}
\label{stimaU}
U(v) \sim \frac{v^2}{\tau_c} \hspace{2cm} \mbox{$v \lesssim \sqrt{T_g}$} \\
U(v) \sim v^{\beta} \hspace{2cm} \mbox{$v \gtrsim \sqrt{T_g}$} \nonumber
\end{eqnarray}
where $\beta=\beta'+1<2$.
It is clear now, looking at eq. (\ref{angelo5}),  
that when $\tau < \tau_c$ (i.e., in the non-clusterized 
regime) the argument of the exponential is dominated by $v^2/\tau$ and 
therefore a gaussian
is expected for $P(v)$ with variance $T_F$. In the opposite regime, when 
$\tau > \tau_c$ the distribution is a gaussian with variance $\frac{\tau_c}
{\tau}T_F$ at low velocities, a simple exponential (if $\beta=1$) at high
 velocities, and a 
gaussian with variance $T_F$ at extremely high velocities, but this very
far tails practically cannot be observable. In figure (\ref{ultima}) 
the tails of the distributions of velocities (from the simulation of the
model of section \ref{modello2}) for three different choices of
parameters are presented: in case (a), when $\tau < \tau_c$, we observe
a Gaussian distribution; in case (b), when $\tau > \tau_c$, we can fit
the tail with the function $\exp(-v^{3/2}/b)$, and this is
in agreement with the analytical calculation performed by van Noije and Ernst
\cite{ernst}; finally in case (c), when $\tau >> \tau_c$ we observe a
simple exponential tail, as we may expect from the argument presented above.

\section{Conclusions and open problems}
\label{fine}
In this paper a class of models of \emph{granular gases} in one and 
two dimensions has been studied by mean of computer simulations and 
analytical investigations. We think to this class of
models as the natural, and more physical, extension of previous models
in the domain of granular kinetics \cite{kadanoff} - \cite{peng}.
 In the models here
proposed, by effect of balance between Brownian driving and 
inelastic collisions, one has a good thermodynamic limit; 
furthermore, these models present a
rich phenomenology as several regimes are 
observed by tuning the physical parameters, that is the time of viscous 
interaction $\tau$ and the coefficient of restitution for inelastic
 collisions $r$. The two extreme behaviors of those models are the
Gaussian/homogeneous regime and the non-Gaussian/clusterized one.
In the homogeneous phase, the system may be described almost as a perfect
gas in equilibrium at a temperature close to that of the external driving (or
a bit lower), showing the absence of densities instabilities and a Maxwellian
distribution of velocities.
The out-of-equilibrium phase, on the other side, presents strong fluctuations
of density (clusters and collapse) with self-similar density distribution and
a stationary fractal dimension, while there is a strong enhancement of
high energy tails in the distribution of velocities. This  dramatic
breaking of the equipartition law has to be kept into account in modeling the
hydrodynamics of granular media. 
Furthermore, we explained the origin of the different degrees of
 clusterization by means of a class of {\em balls-in-the-boxes} models,
showing that the effect of inelasticity may be viewed as a bias to the
transition rates of these random processes: in this context we showed that
the non-Gaussian distribution of velocities is recovered assuming a local
equilibrium with a temperature which depends on the local density.
The non-Gaussian behavior has been also analytically investigated with the 
help of a model in which the effect of collisions is treated as a mean-field
force on each particle and using the fact 
that this force has a different dependence
on the impact velocity whether the particle is in a cluster or outside of it. 
Diffusion of particles has been also investigated in the simulation of models
of section \ref{modello2}: no anomalous diffusion has been
 observed. The diffusion coefficients for the non-Gaussian regime have been
reported in figures (\ref{fig_riep1}) and (\ref{fig_riep2}) showing a
 weak (and apparently non-monotonic) dependency on the restitution coefficient
$r$. A measure of velocity correlation function $<v(t)v(t+\tau)>$, which
appeared not to be a trivial exponential but likely a superposition of
different exponentials (therefore still integrable in time), has convinced us
 that, even in the clusterized regime, the particles forget their previous
velocities rather quickly due to collisions, that is: they enter and exit a
cluster frequently enough to not affect average diffusion; however, in the
clusterized regime the diffusion process is dominated by inter-particles 
collisions, whereas in the homogeneous one the diffusion is dominated by
the Brownian motion imposed by the model. This is only a rough picture, to be
furtherly investigated.

An important task to accomplish should be the research of an equation of state
for this class of gases, useful in an eventual hydrodynamic description of
them. The observed relation between local temperature and density (see
figures (\ref{fig_enfluct1d}), (\ref{fig_ef1}) and (\ref{fig_ef2}) and the
discussion in section \ref{vittorio}) is the
starting point in this project. Analytical expressions of the pressure
have to include the usual streaming term $\rho <v^2>$ (where $\rho$ is the
local density and $<\;>$ is an ensemble average) as well as a collisional term
which is important in the regions where the density is high: the streaming
term, as a consequence of the scaling $<v^2> \sim \rho^{-\beta}$, see the
first of eqs. 
(\ref{inel}), is expected to be proportional to $\sim \rho^{1-\beta}$ if
the picture of local Gaussian equilibrium is confirmed \cite{caglioti}.

{\bf \large Acknowledgments} We thank A. Petri, E. Caglioti
and M. Marsili for useful discussions. V.L. acknowledges financial support 
under project ERBFMBICT961220 and FMRXCT980183.

%******************* BIBLIOGRAFIA *************************************

%*************************** FIGURE ****************************************

\begin{figure}
\centerline{
\psfig{figure=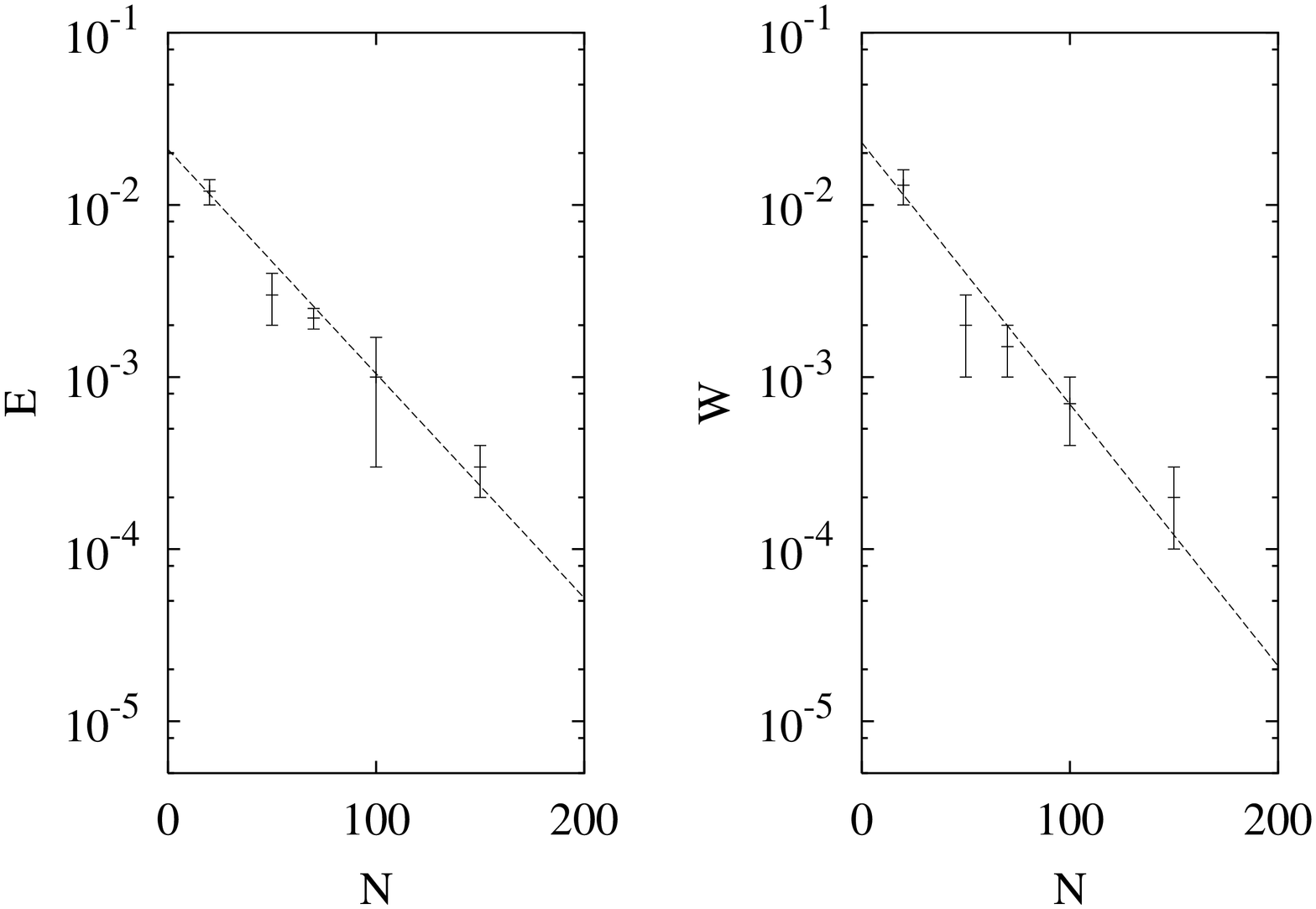,width=8cm,angle=-90}
}
\caption{
Kinetic energy and dissipated energy per particle (as defined in equations 
(\ref{kad_ec}) and (\ref{kad_ed})) vs. N, in Du 
\emph{et al.} [16]
model. Particle density $N/L=100$ is kept constants and 
$r=0.99$.}
\label{fig_nolimit}
\end{figure}

\begin{figure}
\centerline{
\psfig{figure=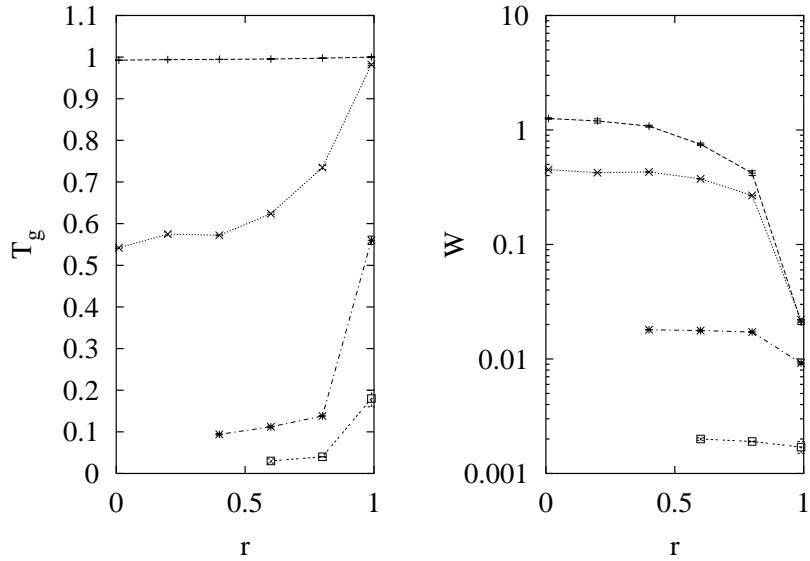,width=8cm,angle=-90}
}
\caption{
The average granular temperature $T_g$ and the average dissipated energy per particle  $W$ 
vs the restitution coefficient $r$ for different values
of $\tau$ and $N=200$. From top to bottom: $\tau=0.01$, $\tau=2$, $\tau=100$,
$\tau=1000$.}
\label{fig_ewriepilog1d}
\end{figure}

\newpage

\begin{figure}
\centerline{
\psfig{figure=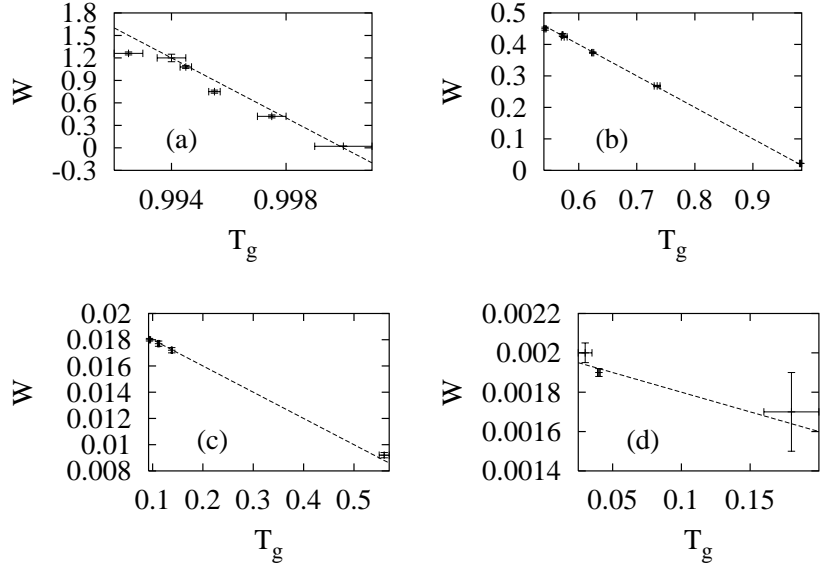,width=8cm,angle=-90}
}
\caption{
Dissipated energy per particle $W$ versus the granular temperature $T_g$, 
for different $\tau$ and different $r$: $\tau=0.01$ (a), $\tau=2$ (b), 
$\tau=100$ (c), $\tau=1000$ (d). The dashed lines represent the relation 
(\ref{relazionewt})}
\label{fig_relazionewt}
\end{figure}

\begin{figure}
\centerline{
\psfig{figure=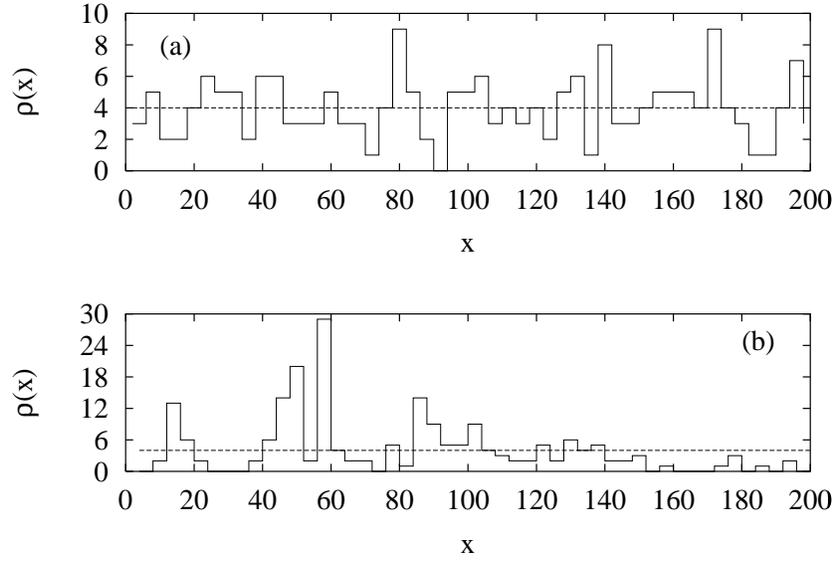,width=8cm,angle=-90}
}
\caption{
Instantaneous density profiles $\rho(x)$ in two regimes: (a)
quasi-equilibrium regime ($\tau=0.01$, $r=0.99$) and (b)
non-equilibrium regime with clusters ($\tau=100$, $r=0.6$). In both
histograms $N=200$ and the dashed horizontal lines represent the
average density, equal to four particles per bin}
\label{fig_density1d}
\end{figure}

\newpage

\begin{figure}
\centerline{
\psfig{figure=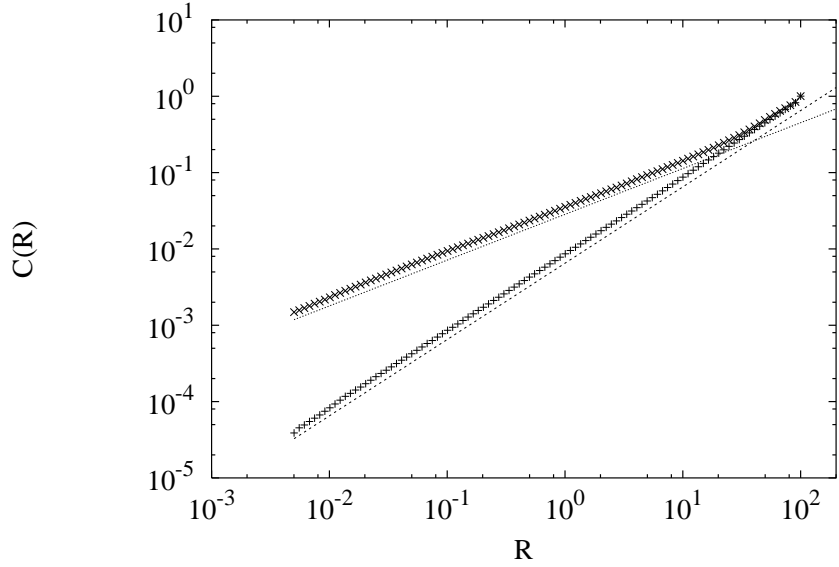,width=8cm,angle=-90}
}
\caption{C(R) vs. R for $\tau=100$, 
$r=0.6$ (top) and $\tau=100$, $r=0.99$ (bottom) with $N=200$. Correlation 
dimension takes 
respectively the values $d_2=0.59$ and $d_2=1$.}
\label{fig_gp1d}
\end{figure}

\begin{figure}
\centerline{
\psfig{figure=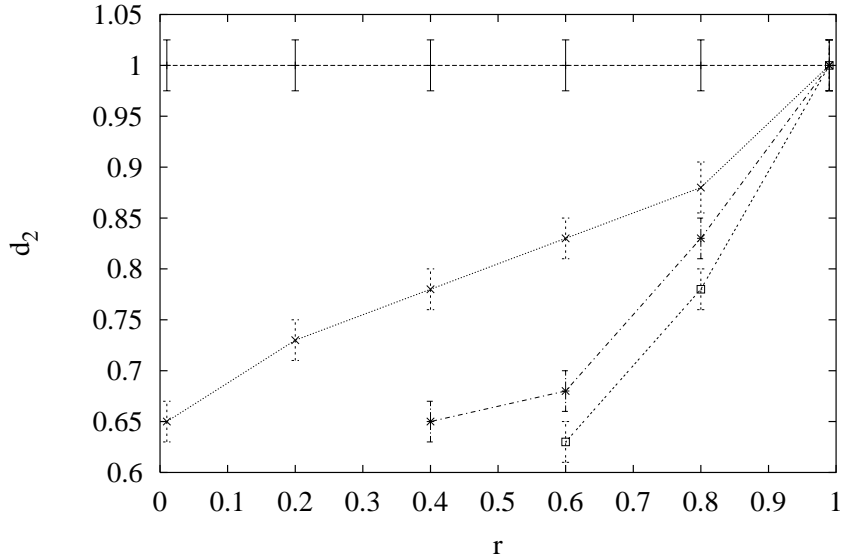,width=8cm,angle=-90}
}
\caption{The correlation dimension $d_2$ vs. $r$ 
for different values of $\tau$: from top to bottom $\tau$ is 0.01, 2, 100 
and 1000. $N=200$
in all simulations}
\label{fig_gp1dtot}
\end{figure}

\newpage

\begin{figure}
\centerline{
\psfig{figure=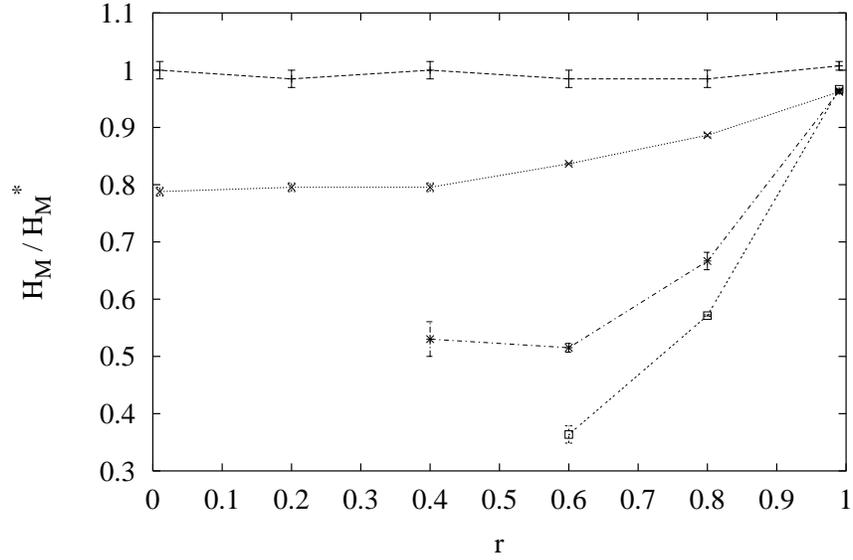,width=8cm,angle=-90}
}
\caption{
$H_M/H_M^*$ vs. $r$ for different 
$\tau$: form top to bottom $\tau$ is 0.01, 2, 100 and 1000, 
with N=200 and M=80 ($H_M^* \approx 63$)}
\label{fig_entropy1d}
\end{figure}

\begin{figure}
\centerline{
\psfig{figure=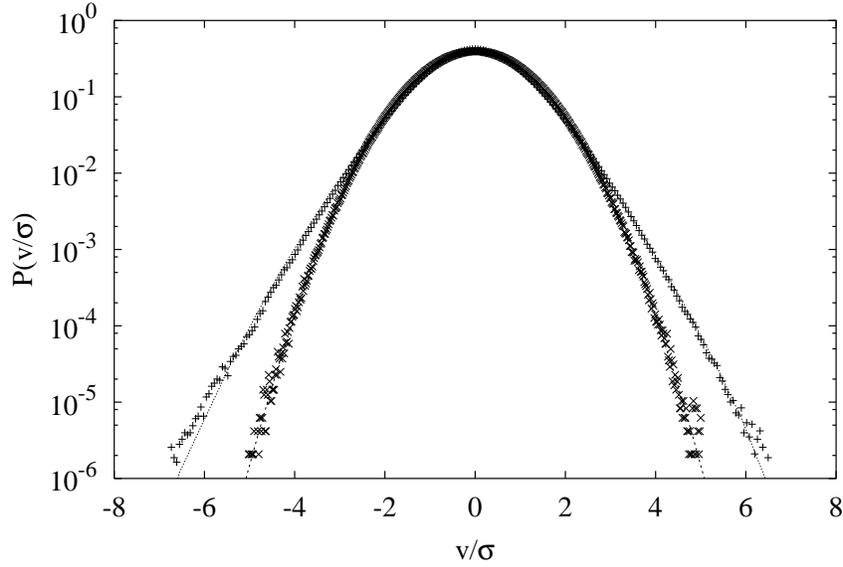,width=8cm,angle=-90}
}
\caption{The rescaled velocity distribution $P(v/\sigma)$ vs. $v/\sigma$. 
Pluses (+) are data from simulation with $\tau=100$, $r=0.7$. Crosses are data
with $\tau=0.01$,$r=0.99$. The dot-dashed line represents the gaussian 
distribution, while the dashed line represents the fit discussed in 
section \ref{vittorio}.}
\label{fig_distvel1d}
\end{figure}

\newpage

\begin{figure}
\centerline{
\psfig{figure=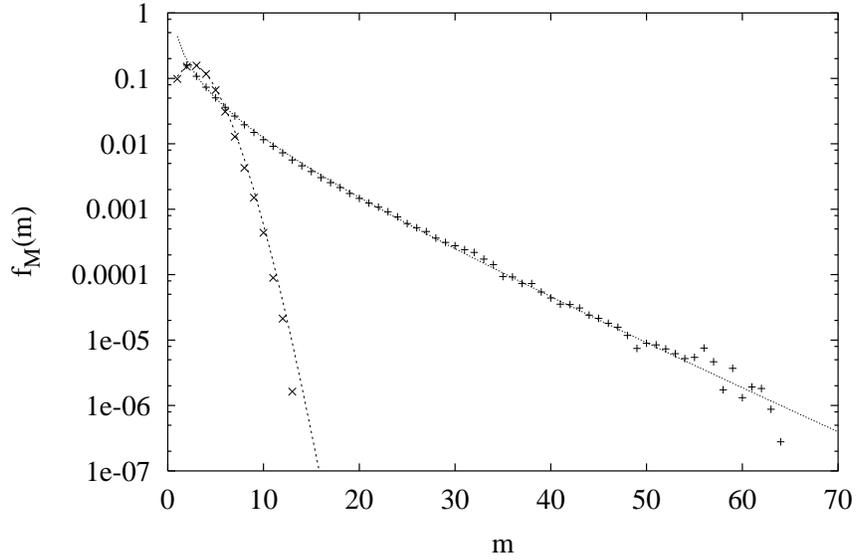,width=8cm,angle=-90}
}
\caption{
Density distribution vs. $m$ for two different choices of parameters: $\tau=0.01$, 
$r=0.99$ and $\tau=100$, $r=0.7$, $N=300$ and $M=100$. The former
is fitted by a Poisson distribution with $\lambda=3$, while the latter is
fitted by $\frac{1}{m}e^{-0.14*m}$}
\label{fig_denfluct1d}
\end{figure}

\begin{figure}
\centerline{
\psfig{figure=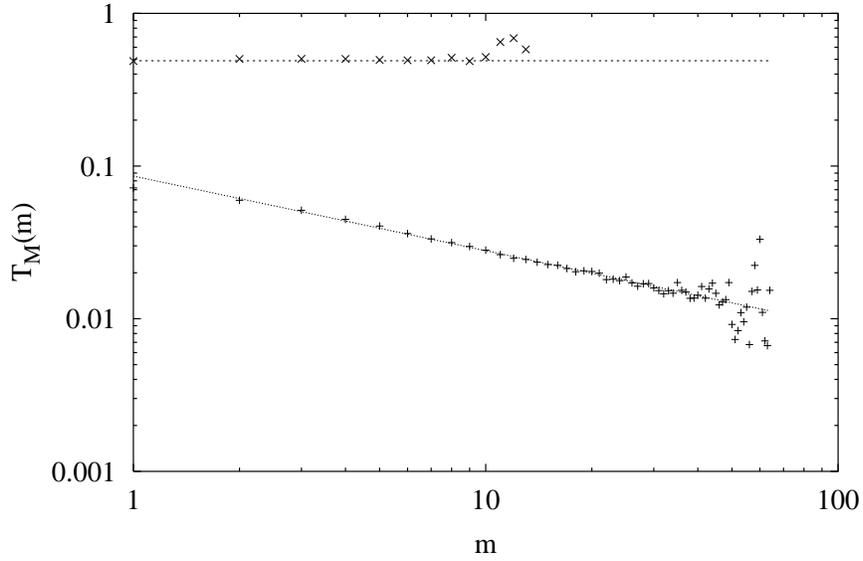,width=8cm,angle=-90}
}
\caption{Box granular temperature $T_M(m)$ against $m$, when $\tau=0.01$, 
$r=0.99$ and $\tau=100$, $r=0.7$ (in this case th fit $m^{-0.5}$ is plotted).}
\label{fig_enfluct1d}
\end{figure}

\newpage

\begin{figure}
\centerline{
\psfig{figure=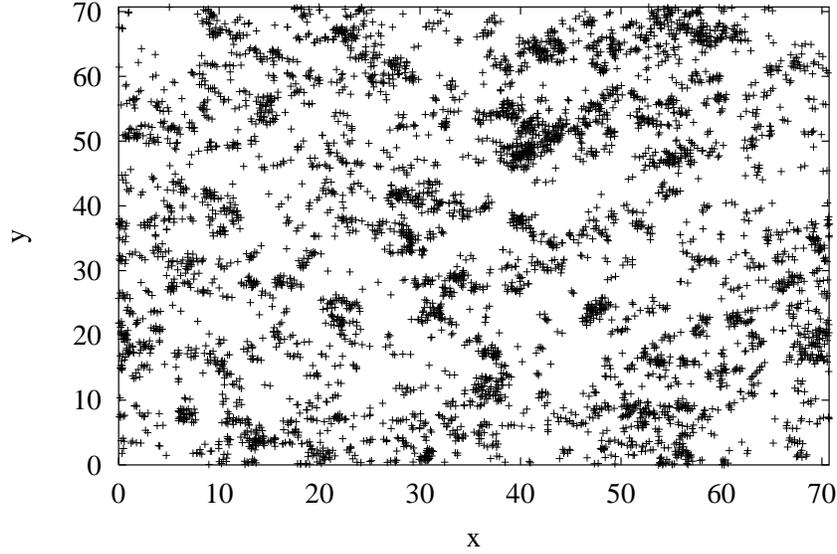,width=8cm,angle=-90}
}
\caption{Snapshot of particle distribution in 2 dimensions in the clusterized
regime. $N=5000$, $\tau_c=0.5$, $l=0.63$, $\tau=100$, $r=0.01$.} 
\label{fig_de2}
\end{figure}

\begin{figure}
\centerline{
\psfig{figure=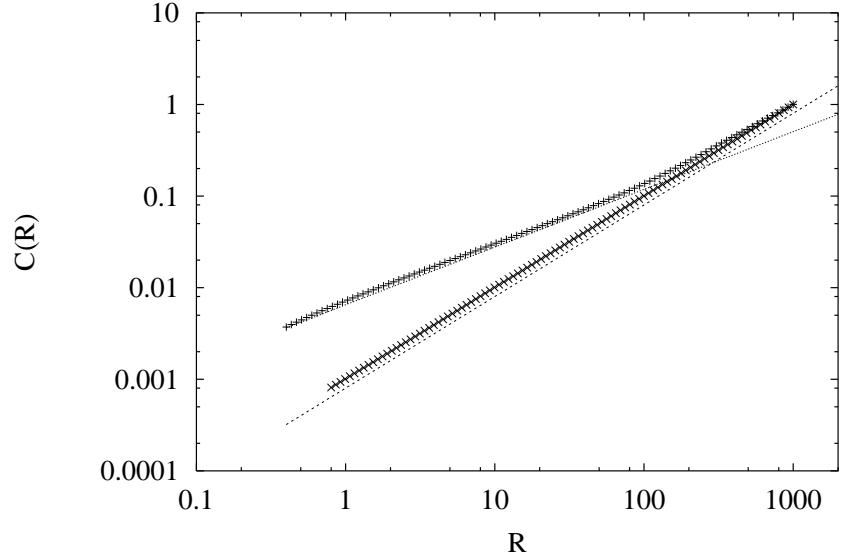,width=8cm,angle=-90}
}
\caption{C(R) vs. R in the one-dimensional system. $N=2000$, $\tau_c=0.5$,
$l=0.4$. The top curve is for $\tau=100$, $r=0.5$, the bottom one is for
$\tau=0.01$, $r=0.99$. Correlation dimension is, respectively, $d_2=0.55$ and
$d_2=1$.}
\label{fig_fr1}
\end{figure}

\newpage

\begin{figure}
\centerline{
\psfig{figure=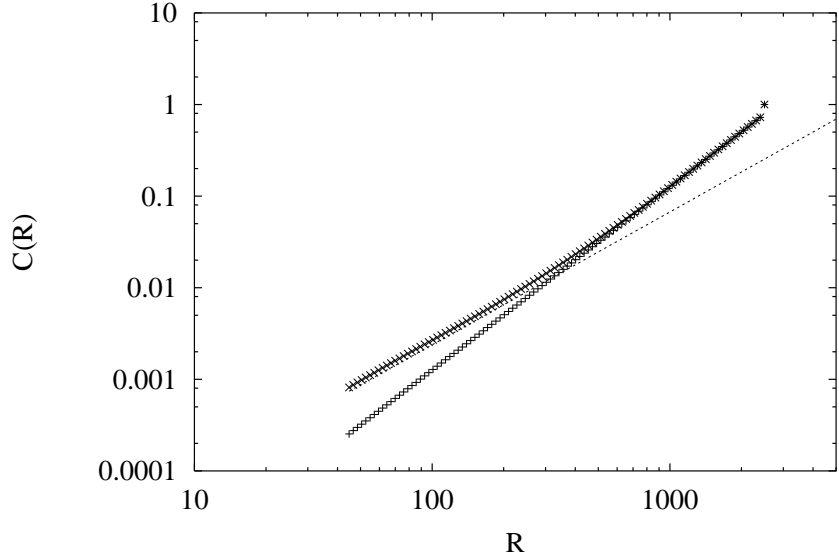,width=8cm,angle=-90}
}
\caption{C(R) vs. R in 2-d. $N=5000$, $\tau_c=0.5$, $l=0.71$. Top curve is for
$\tau=100$, $r=0.5$, while the bottom one is for $\tau=0.01$,
$r=0.99$. The correlation
dimension is $d_2=1.45$  and 
$d_2=2$, respectively}
\label{fig_fr2}
\end{figure}

\begin{figure}
\centerline{
\psfig{figure=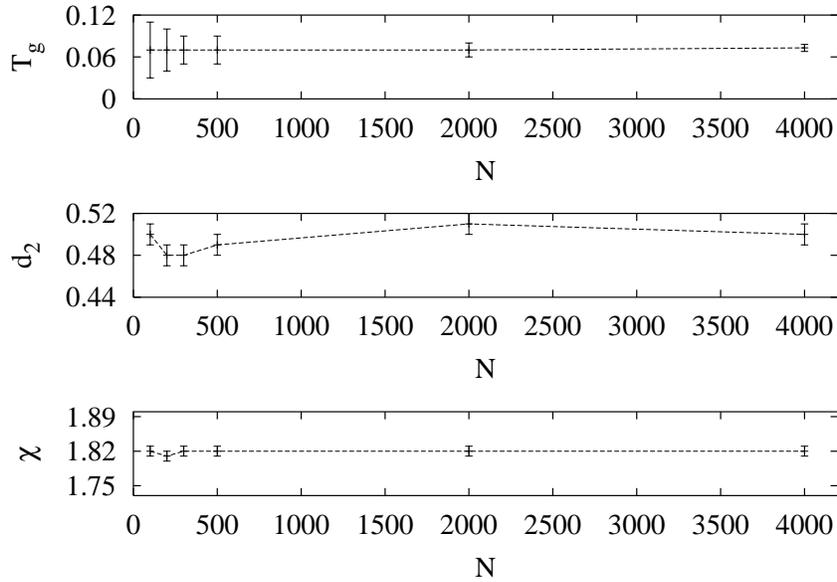,width=8cm,angle=-90}
}
\caption{The granular temperature $T_g$, fractal dimension $d_2$ and 
collision rate $\chi$ vs number of particle $N$, for the model in one
dimension, with $\tau_c=0.5$, $l=0.4$, $\tau=100$ and $r=0.5$.} 
\label{fig_lt1}
\end{figure}

\newpage

\begin{figure}
\centerline{
\psfig{figure=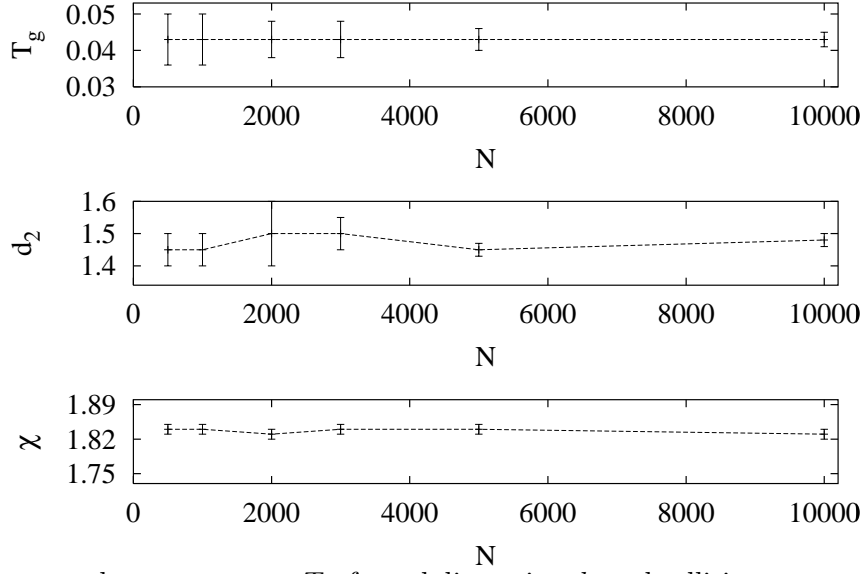,width=8cm,angle=-90}
}
\caption{The granular temperature $T_g$, fractal dimension $d_2$ and
collision rate $\chi$ against number of particle $N$, for the model in two
dimensions, with $\tau_c=0.5$, $l=0.63$, $\tau=100$, $r=0.5$}
\label{fig_lt2}
\end{figure}

\begin{figure}
\centerline{
\psfig{figure=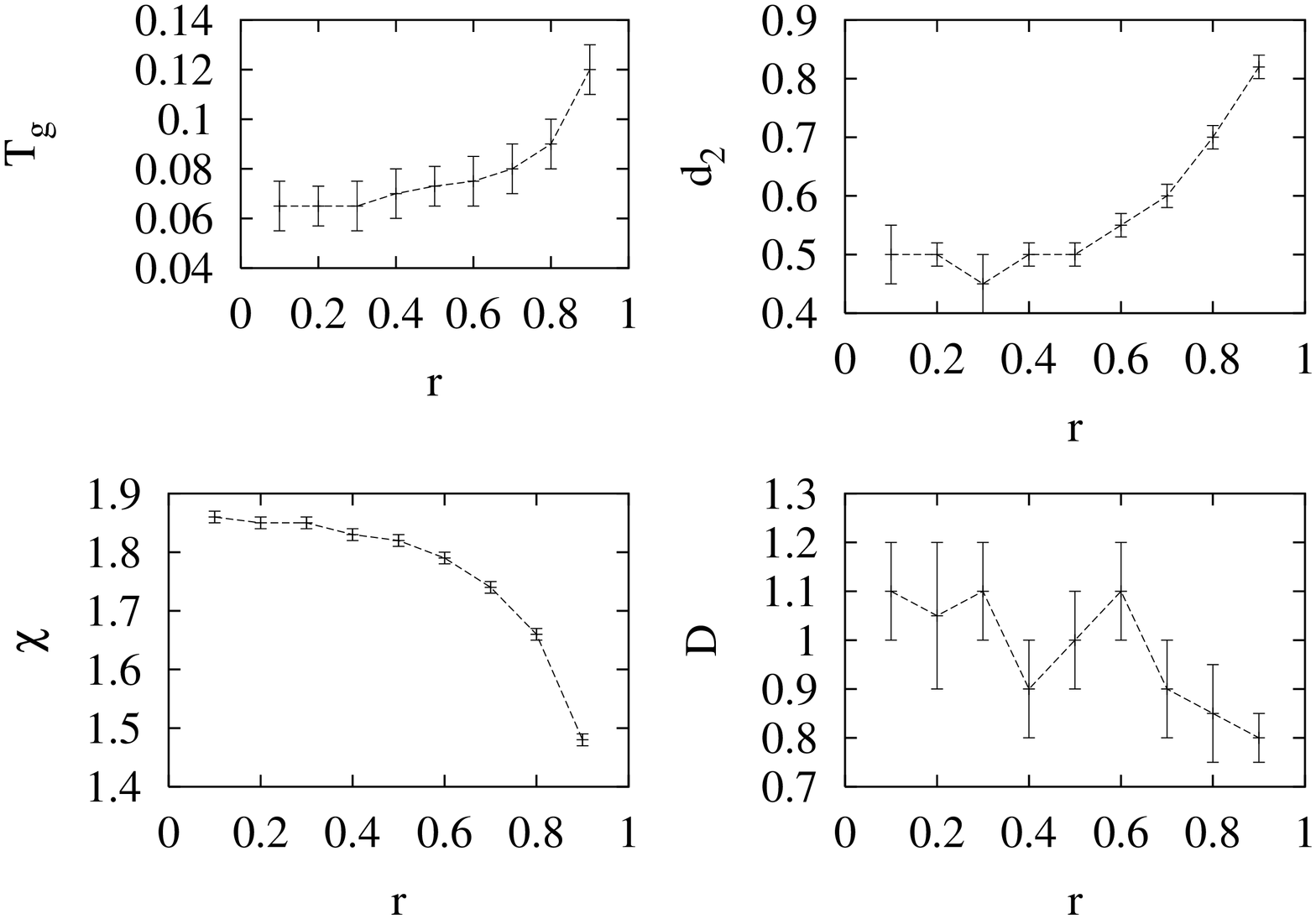,width=8cm,angle=-90}
}
\caption{The granular temperature $T_g$, fractal dimension $d_2$, 
collision rate $\chi$ and diffusion coefficient $D$ against restitution 
coefficient $r$,  for the model in one
dimension, with $N=4000$, $\tau_c=0.5$, $l=0.4$, $\tau=100$ and $r=0.5$.} 
\label{fig_riep1}
\end{figure}

\newpage

\begin{figure}
\centerline{
\psfig{figure=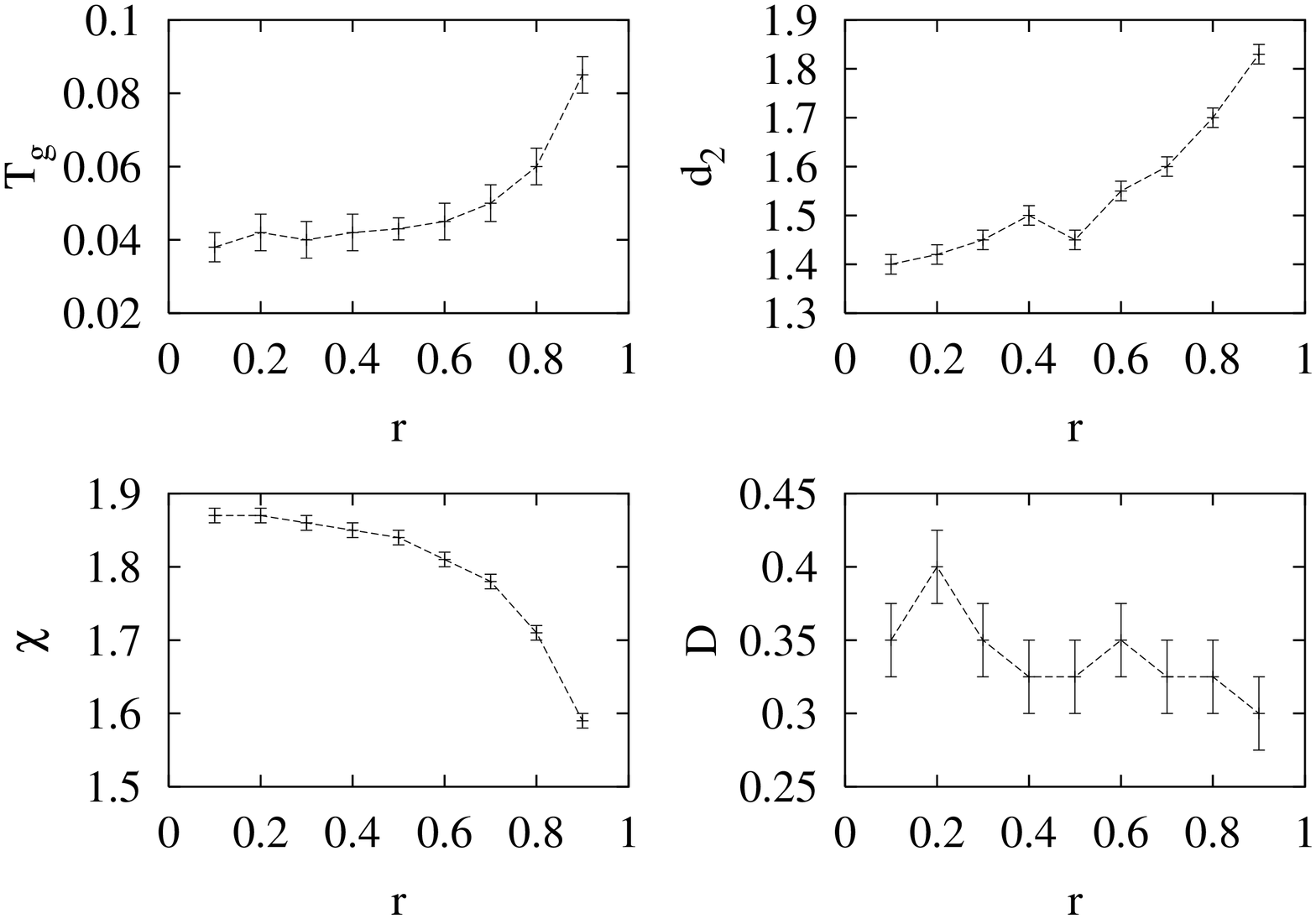,width=8cm,angle=-90}
}
\caption{The granular temperature $T_g$, fractal dimension $d_2$,
collision rate $\chi$ and diffusion coefficient $D$ against restitution
coefficient $r$, for the model in two
dimensions, with $N=3000$, $\tau_c=0.5$, $l=0.63$, $\tau=100$, $r=0.5$}
\label{fig_riep2}
\end{figure}

\begin{figure}
\centerline{
\psfig{figure=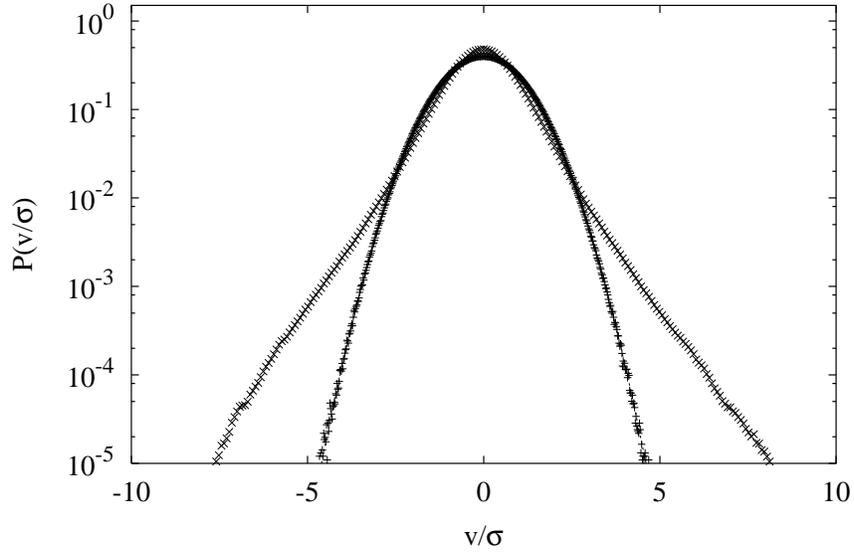,width=8cm,angle=-90}
}
\caption{Distribution of velocities in a gaussian ($\tau=0.01$, $r=0.99$) 
and a non-gaussian regime ($\tau=100$, $r=0.5$) for the one dimensional 
system. In both cases $N=2000$, 
$\tau_c=0.5$, $l=0.4$. The dashed curve represents the gaussian.}
\label{fig_ve1}
\end{figure}

\newpage

\begin{figure}
\centerline{
\psfig{figure=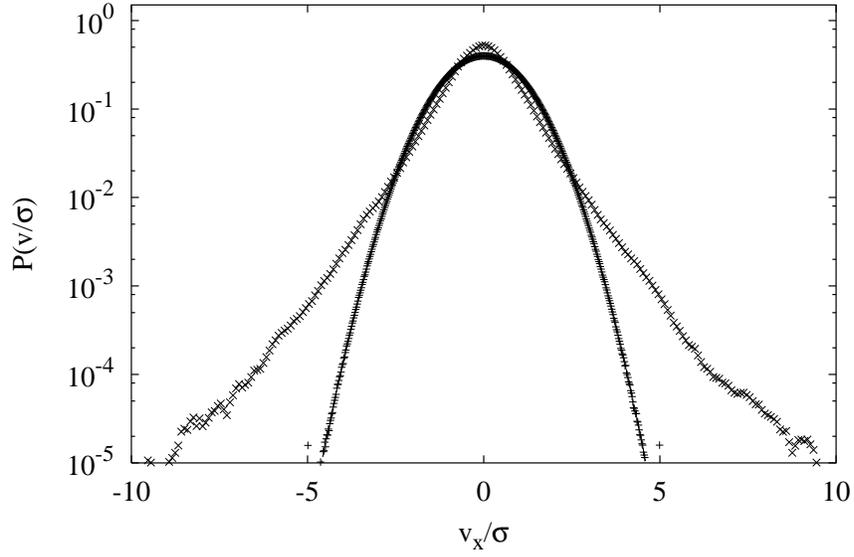,width=8cm,angle=-90}
}
\caption{Distribution of velocities in a gaussian ($\tau=0.01$, $r=0.99$) and
a non-gaussian regime ($\tau=100$, $r=0.5$) for the two dimensions case.
It is always $N=10000$, $\tau_c=0.05$, $l=0.22$ and the dashed line represents
the gaussian}
\label{fig_ve2}
\end{figure}

\begin{figure}
\centerline{
\psfig{figure=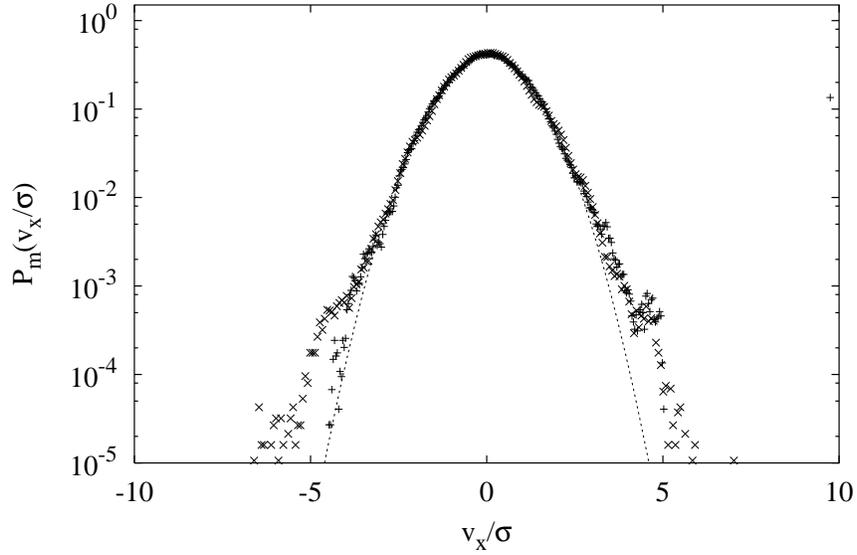,width=8cm,angle=-90}
}
\caption{Distribution of velocities restricted to number density $m=1$ (pluses
'+') and $m=5$ (crosses 'X'), in a 2-dimensional case, with $N=10000$, 
$\tau_c=0.05$, $l=0.22$, $\tau=100$, $r=0.5$.}
\label{fig_ve_de}
\end{figure}

\newpage

\begin{figure}
\centerline{
\psfig{figure=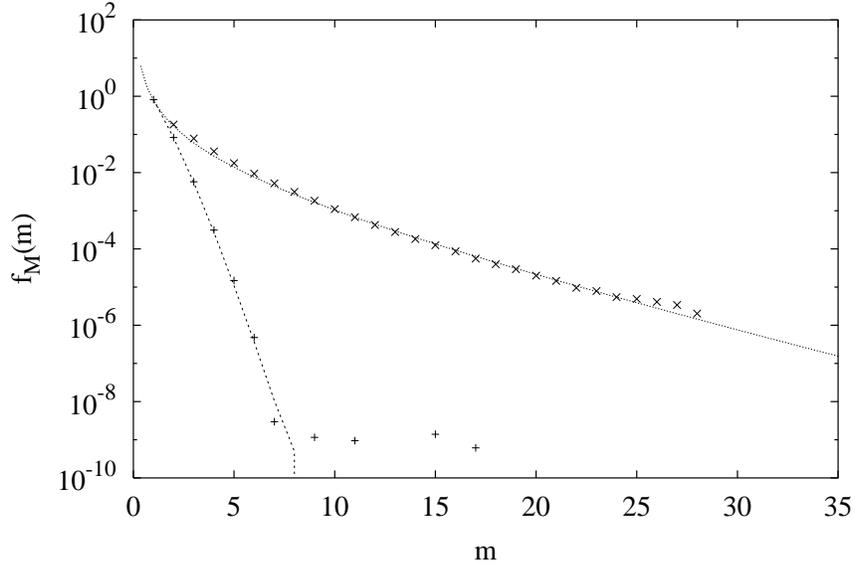,width=8cm,angle=-90}
}
\caption{Distribution density $f_M(m)$ vs. $m$ for two one-dimensional cases: 
$\tau=100$, $r=0.5$ and $\tau=0.01$, $r=0.99$. In both cases: $N=500$,
$\tau_c=0.5$, $l=0.4$, $M=12000$. There are
two curves superimposed: a Poisson function (with $\lambda=N/M \approx 0.04$)
and $m^{-1.95} \exp(-0.26*m)$ fit for the clusterized regime.}
\label{fig_df1}
\end{figure}

\begin{figure}
\centerline{
\psfig{figure=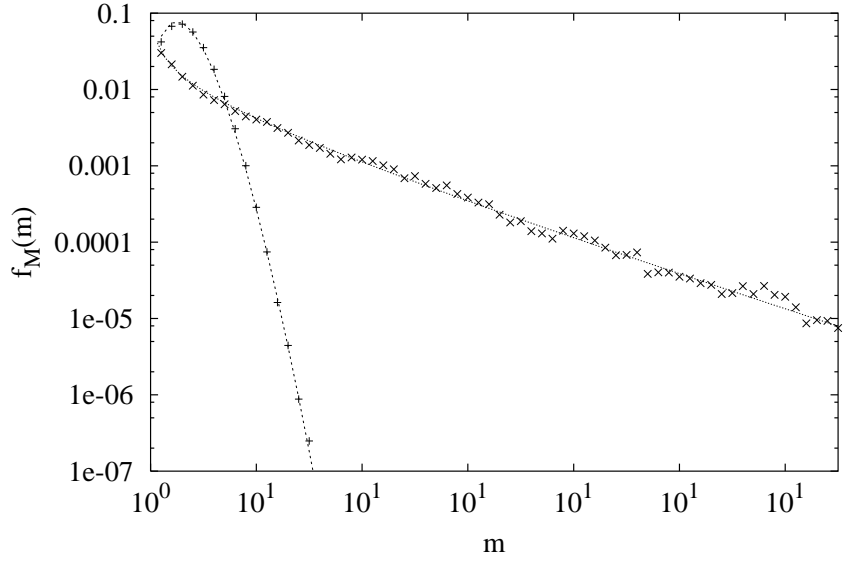,width=8cm,angle=-90}
}
\caption{Distribution density $f_M(m)$ vs. $m$ for two bidimensional cases: 
$\tau=100$, $r=0.5$ and $\tau=0.01$, $r=0.99$. In both cases: $N=10000$,
$\tau_c=0.05$, $l=0.22$, $M=3200$. There are
two curves superimposed: a Poisson function (with $\lambda=N/M=3.125$)
and $m^{-0.5} \exp(-0.097*m)$ fit for the clusterized regime.}
\label{fig_df2}
\end{figure}

\newpage

\begin{figure}
\centerline{
\psfig{figure=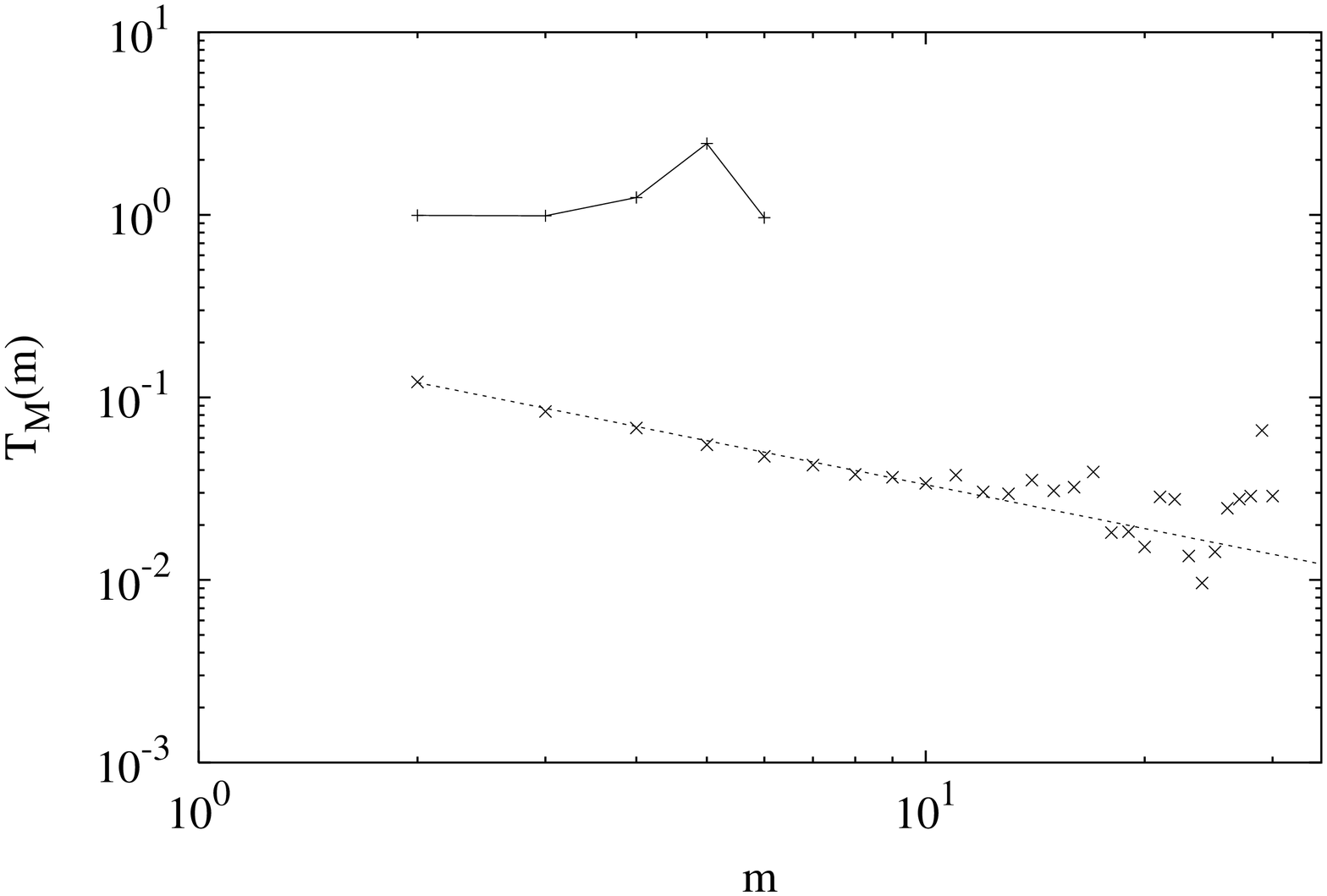,width=8cm,angle=-90}
}
\caption{Box granular temperature $T_M(m)$ vs. $m$ for two one-dimensional cases: 
$\tau=100$, $r=0.5$ and $\tau=0.01$, $r=0.99$. In both cases: $N=500$,
$\tau_c=0.5$, $l=0.4$, $M=12000$. The gaussian case is constant, while
the non-gaussian case is fitted by $\sim m^{-0.8}$.}
\label{fig_ef1}
\end{figure}

\begin{figure}
\centerline{
\psfig{figure=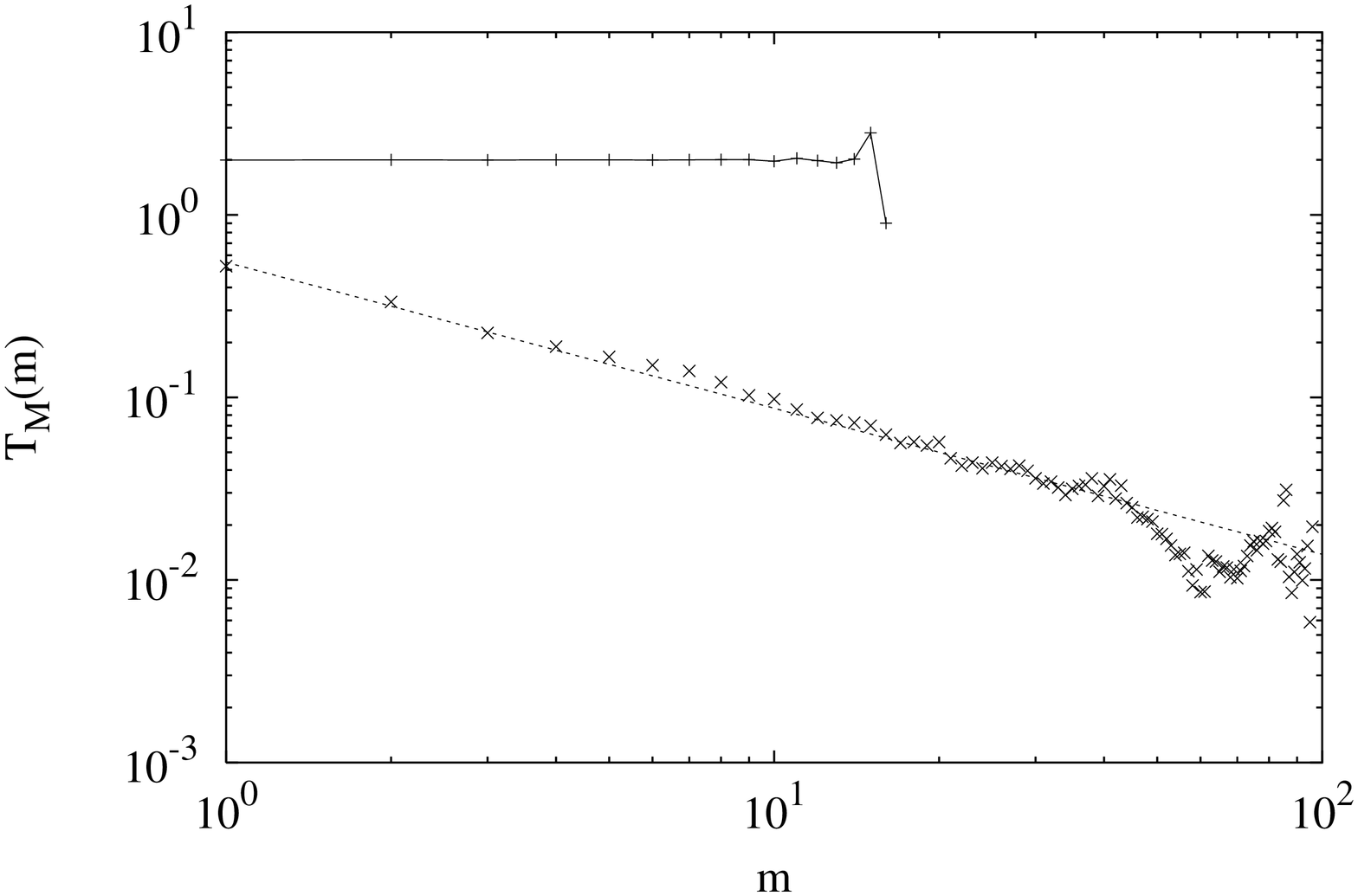,width=8cm,angle=-90}
}
\caption{Box granular temperature $T_M(m)$ vs. $m$ for two bidimensional cases: 
$\tau=100$, $r=0.5$ and $\tau=0.01$, $r=0.99$. In both cases: $N=10000$,
$\tau_c=0.05$, $l=0.22$, $M=3200$. The gaussian case is constant, while
the non-gaussian one is fitted by $\sim m^{-0.8}$.}
\label{fig_ef2}
\end{figure}

\newpage

\begin{figure}
\centerline{
\psfig{figure=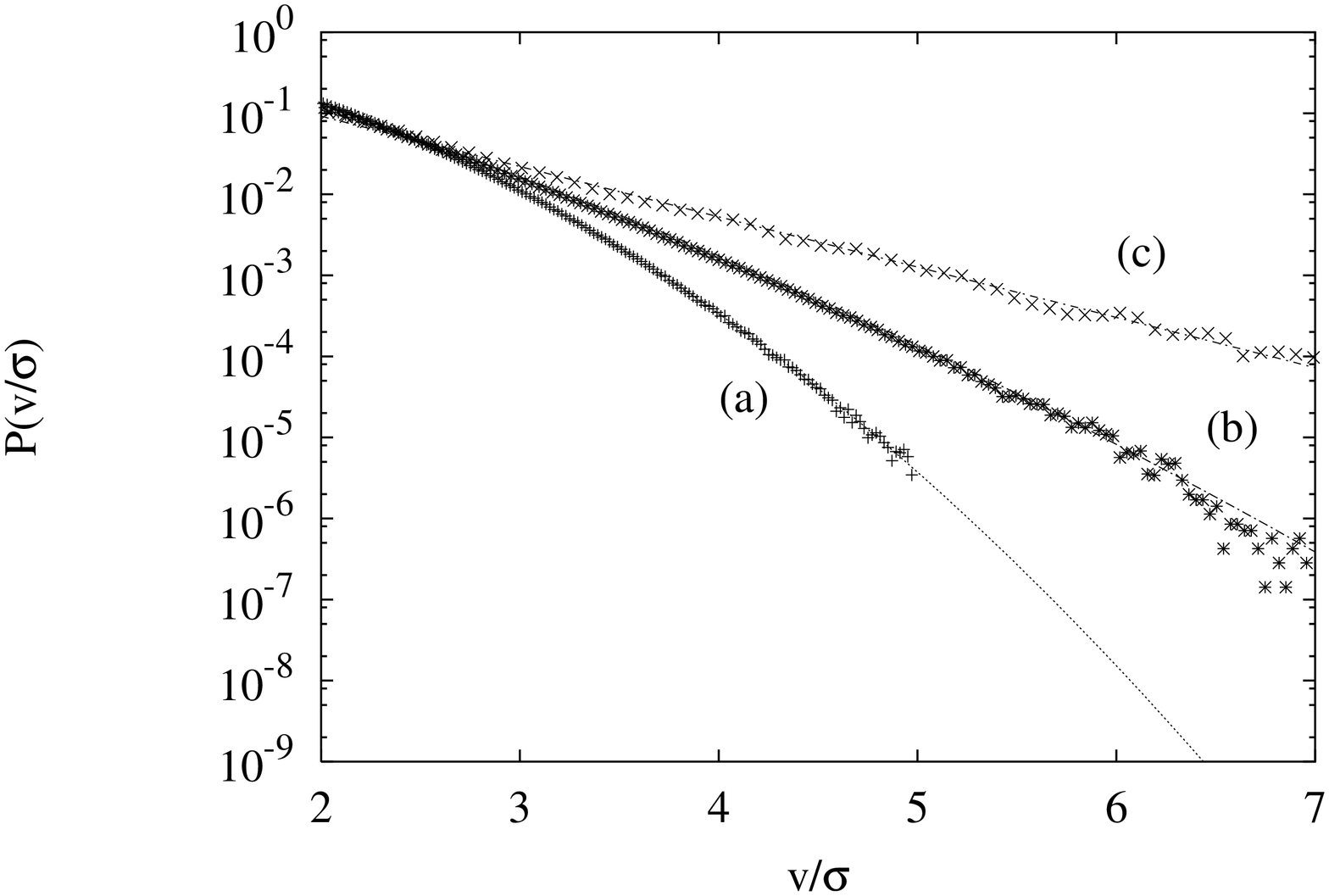,width=8cm,angle=-90}
}
\caption{Rescaled distributions of velocities (particular) for three different
 choices of parameters, in two dimensions: (a) $N=10000$, $\tau=0.01$, 
$r=0.99$, with Gaussian 
fit; (b) $N=3000$, $\tau=5$, $r=0.5$ with the fit $\sim \exp(-v^{3/2}/1.25)$;
(c) $N=10000$, $\tau=100$, $r=0.2$ with the fit $\sim \exp(-v/0.7)$. In the
cases (a) and (c): $\tau_c=0.05$, $l=0.22$. In the case (b): $\tau_c=0.5$
and $l=0.63$.}
\label{ultima}
\end{figure}


\begin{references}
\bibitem{cassi} We are indebted with Davide Cassi that recalled
us the solution of this combinatorial problem.
\end{references}
\end{document}